\documentclass[twocolumn]{aastex63}  
\usepackage{color}
\usepackage{graphicx}
\usepackage{amsmath}
\usepackage{amssymb}	
\usepackage{diagbox}
\usepackage{threeparttable}

\usepackage{ulem}
\usepackage{natbib}
\usepackage{float}
\usepackage{longtable, threeparttablex, booktabs, url}
\usepackage{verbatim}
\usepackage{afterpage}
\usepackage{enumitem}
\usepackage{hyperref}

\shorttitle{GBT Observations of FRB 20240114A}
\shortauthors{Xie et al.}

\let\oldequation\equation
\let\oldendequation\endequation

\renewenvironment{equation}
  {\linenomathNonumbers\oldequation}
  {\oldendequation\endlinenomath}

\providecommand{\DIFaddbegin}{} 
\providecommand{\DIFaddend}{} 
\providecommand{\DIFdelend}{} 

\begin{document}

\title{Polarization Characteristics of the Hyperactive FRB~20240114A}

\correspondingauthor{Yi Feng; Di Li; Yong-Kun Zhang}
\email{yifeng@zhejianglab.org; dili@tsinghua.edu.cn;\\ ykzhang@nao.cas.cn}

\author[0000-0001-5649-2591]{Jin-Tao Xie}
\affiliation{School of Computer Science and Engineering, Sichuan University of Science and Engineering, Yibin 644000, China}

\author[0000-0002-0475-7479]{Yi Feng}
\affiliation{Research Center for Astronomical Computing, Zhejiang Laboratory, Hangzhou 311100, China}
\affil{Institute for Astronomy, School of Physics, Zhejiang University, Hangzhou 310027, China}

\author[0000-0003-3010-7661]{Di Li}
\affiliation{New Cornerstone Science Laboratory, Department of Astronomy, Tsinghua University, Beijing 100084, China}
\affiliation{National Astronomical Observatories, Chinese Academy of Sciences, A20 Datun Road, Chaoyang District, Beijing 100101, China}
\affiliation{Zhejiang Lab, Hangzhou, Zhejiang 311121, People’s Republic of China}

\author[0000-0002-8744-3546]{Yong-Kun Zhang}
\affiliation{National Astronomical Observatories, Chinese Academy of Sciences, Beijing 100101, China}
\affiliation{University of Chinese Academy of Sciences, Beijing 100049, China}

\author{Deng-Ke Zhou}
\affiliation{Research Center for Astronomical Computing, Zhejiang Laboratory, Hangzhou 311100, China}

\author[0000-0003-4721-4869]{Yuanhong Qu}
\affil{Nevada Center for Astrophysics, University of Nevada, Las Vegas, NV 89154, USA}
\affil{Department of Physics and Astronomy, University of Nevada, Las Vegas, NV 89154, USA}

\author[0000-0002-6165-0977]{Xiang-han Cui}
\affiliation{National Astronomical Observatories, Chinese Academy of Sciences, Beijing 100101, China}
\affiliation{International Centre for Radio Astronomy Research, Curtin Institute of Radio Astronomy, Perth 6102, Australia}
\affiliation{University of Chinese Academy of Sciences, Beijing 100049, China}

\author{Jian-Hua Fang}
\affiliation{Research Center for Astronomical Computing, Zhejiang Laboratory, Hangzhou 311100, China}

\author{Jia-Ying Xu}
\affiliation{Research Center for Astronomical Computing, Zhejiang Laboratory, Hangzhou 311100, China}

\author{Chen-Chen Miao}
\affiliation{Research Center for Astronomical Computing, Zhejiang Laboratory, Hangzhou 311100, China}

\author{Mao Yuan}
\affiliation{National Space Science Center, Chinese Academy of Sciences, Beijing 100101, China}

\author[0000-0002-9390-9672]{Chao-Wei Tsai}
\affil{National Astronomical Observatories, Chinese Academy of Sciences, Beijing 100101, China}
\affiliation{Institute for Frontiers in Astronomy and Astrophysics, Beijing Normal University,  Beijing 102206, China}
\affiliation{University of Chinese Academy of Sciences, Beijing 100049, China}

\author{Pei Wang}
\affil{National Astronomical Observatories, Chinese Academy of Sciences, Beijing 100101, China}
\affil{Institute for Frontiers in Astronomy and Astrophysics, Beijing Normal University, Beijing 102206, China}

\author{Chen-Hui Niu}
\affil{Institute of Astrophysics, Central China Normal University, Wuhan 430079, Hubei, China}

\author[0000-0001-5738-9625]{Xianglei Chen}
\affil{National Astronomical Observatories, Chinese Academy of Sciences, Beijing 100101, China}

\author{Mengyao Xue}
\affil{National Astronomical Observatories, Chinese Academy of Sciences, Beijing 100101, China}

\author{Junshuo Zhang}
\affil{National Astronomical Observatories, Chinese Academy of Sciences, Beijing 100101, China}
\affiliation{University of Chinese Academy of Sciences, Beijing 100049, China}

\begin{abstract}
Fast radio bursts (FRBs) are transient radio bursts of extragalactic origin characterized by millisecond durations and high luminosities. We report on observations of FRB 20240114A conducted with the Robert C. Byrd Green Bank Telescope (GBT) at frequencies ranging from 720 to 920 MHz. A total of 437 bursts were detected, with a single observation recording 365 bursts over 1.38 hours, corresponding to a burst rate of  264 bursts per hour. The average rotation measures (RMs) were {\rm $347.0 \pm 1.0$} rad m$^{-2}$ on February 23, 2024, and {\rm $353.7 \pm 0.6$} rad m$^{-2}$ on March 1, 2024. Of the 301 bursts with detected RMs, 81\% have a linear polarization fraction greater than 90\%, and 14\% exhibit circular polarization with a signal-to-noise ratio $> 5$. Our sample also displayed polarization angle swings. We compared the linear polarization fraction of FRB~20240114A with those of the repeating sources FRB~20201124A and FRB~20220912A. Our analysis reveals that all three exhibit similar distributions in both linear and circular polarization fractions. These results indicate that the three sources share the same radiation mechanism. We analyze the fluence and waiting-time distributions of FRB~20240114A, revealing a right-skewed fluence distribution and a bimodal waiting-time structure, suggesting intrinsic emission timescales and potential multiple burst populations. Additionally, We present a novel method to determine the frequency range of bursts based on their spectral characteristics. This algorithm is independent of spectral models and remains unaffected by the removal of interference-affected channels in the data, ensuring robust performance.

\end{abstract}


\section{Introduction} \label{sec:intro}

Fast radio bursts (FRBs) are bright radio bursts originating at cosmological distances first discovered by \cite{2007Sci...318..777L}. Their cosmological origins and high energies make them valuable tools for studying the cosmic web, Galactic halos, and baryons (\citealt{2016Sci...354.1249R}, \citealt{2019Sci...366..231P}, \citealt{2020Natur.581..391M}). However, the progenitors and radiation mechanisms of FRBs remain unclear. A particularly intriguing class of FRBs is the repeating FRBs, which emit bursts multiple times. To date, more than 800 FRB sources have been discovered, of which only 66 are repeating FRBs, accounting for approximately 8.1\% of the total \footnote{https://blinkverse.alkaidos.cn/} \citep{2023Univ....9..330X}.

Large datasets of repeating FRBs are crucial for uncovering their properties. For example, 1,652 bursts from FRB~20121102A reveal that the burst energy distribution is bimodal and well-characterized by a combination of a log-normal function and a generalized Cauchy function \citep{2021Natur.598..267L}. Similarly, 1,863 bursts from FRB~20201124A reveal irregular short-term variations in the rotation measure (RM), enabling the systematic study of its polarization properties \citep{2022Natur.609..685X}. Additionally, 1,076 bursts from FRB~20220912A allow for precise measurements of its cumulative energy distribution, spectral index of the synthetic spectrum, and polarization properties \citep{2023ApJ...955..142Z}.
A high burst rate is essential for obtaining a large number of bursts. For instance, FRB~20121102A and FRB~20201124A reached peak burst rates of 122 hr$^{-1}$ \citep{2021Natur.598..267L} and 542 hr$^{-1}$ \citep{2022RAA....22l4002Z}, respectively.

Recently, a high burst rate of FRB~20240114A was reported by the Canadian Hydrogen Intensity Mapping Experiment Fast Radio Burst (CHIME/FRB) collaboration \citep{2024ATel16420....1S}. According to the CHIME report, the dispersion measure (DM) value of FRB~20240114A is approximately 527.7 $\mathrm{pccm^{-3}}$. The maximum Galactic DM contribution is about 38.82 and 49.67 $\mathrm{pccm^{-3}}$ for the YMW16 model \citep{2017ApJ...835...29Y} and the NE2001 model \citep{2002astro.ph..7156C}, respectively. This repeating FRB was subsequently followed up by telescopes worldwide, including Parkes/Murriyang \citep{2024ATel16430....1U, 2024ATel16431....1U}, the Westerbork RT1 25-m telescope \citep{2024ATel16432....1O}, the Five-hundred-meter Aperture Spherical radio Telescope (FAST) \citep{2024ATel16433....1Z, 2024ATel16435....1Z,2024ATel16505....1Z}, the Northern Cross (NC) radio telescope \citep{2024ATel16434....1P, 2024ATel16547....1P}, MeerKAT \citep{2024ATel16446....1T}, the upgraded Giant Metrewave Radio Telescope (uGMRT) \citep{2024ATel16452....1K, 2024ATel16494....1P}, European Very Long Baseline Interferometry (EVN) PRECISE \citep{2024ATel16542....1S}, the five small European radio telescopes \citep{2024ATel16565....1O}, the Nançay Radio Telescope (NRT) \citep{2024ATel16597....1H}, the Allen Telescope Array (ATA) \citep{2024ATel16599....1J}, and the Effelsberg 100-m Telescope \citep{2024ATel16620....1L}.
MeerKAT localized FRB~20240114A to RA = 21h27m39.86s, Dec = +04d19m45.01s (J2000), with a positional uncertainty of 1.4 arcseconds, spatially coincident with the galaxy J212739.84+041945.8 \citep{2024MNRAS.533.3174T}. Subsequent spectral observations using the Optical System for Imaging and low-Intermediate-Resolution Integrated Spectroscopy (OSIRIS) spectrograph on the Gran Telescopio Canarias (GTC) determined the host galaxy of this FRB to have a redshift of z = 0.1300 $\pm$ 0.0002 \citep{2024ATel16613....1B}. The host galaxy is classified as a dwarf star-forming galaxy and makes a significant contribution to the DM \citep{2024ATel16613....1B}.

In this paper, we confirm the hyperactivity of FRB~20240114A, report its polarization properties, and compare them with those of FRB~20201124A and FRB~20220912A. This paper is structured as follows: Section 2 details the observational setup and data acquisition methods employed with the GBT. Section 3 presents the results and discussion of the burst properties and polarization characteristics derived from our observations. Finally, Section 4 summarizes our conclusions.

\section{Observations and Data Processing} \label{sec:Observation}

FRB~20240114A was observed using the GBT's P-Band receiver and the VEGAS digital backend\citep{2015ursi.confE...4P}, covering a frequency range of 720–920 MHz.
We conducted two observation sessions with the GBT on February 23, 2024, and March 1, 2024. Each observation session lasted 2 hours, with approximately 1.33 hours of on-source time after setup and calibration. The data were coherently dedispersed at a DM of 527.7 ,pc~cm$^{-3}$ and recorded as 8-bit samples in polarimetric search mode, following the \textsc{psrfits} standard format \citep{2004PASA...21..302H}. Full-Stokes spectra were captured every 40.96 $\mu$s with a channel width of 0.39 MHz.

We used DRAFTS\footnote{\url{https://github.com/SukiYume/DRAFTS}}, a deep-learning-based single-pulse search tool, for burst searches \citep{DRAFTS}. We first de-dispersed the raw observational data with ${\rm DM}=527.7~\rm pc~cm^{-3}$, downsampled the data by a factor of 4, and then split it into segments of 512 samples each. Each segment was resized to $512 \times 512$ pixels and fed into DRAFTS' classification model for inference. The classification model, based on ResNet \citep{he2016deep} and trained on real FAST data, takes $512 \times 512$ time-frequency data as input and outputs the probability that the segment contains an FRB burst. We selected data segments where the model predicted a probability of over 50\% for containing an FRB burst for further manual inspection. Low signal-to-noise ratio signals were excluded, and the specific times of the remaining bursts in the data were identified.

\begin{figure*}[!htp]
  \centering
  \includegraphics[width=0.49\textwidth]{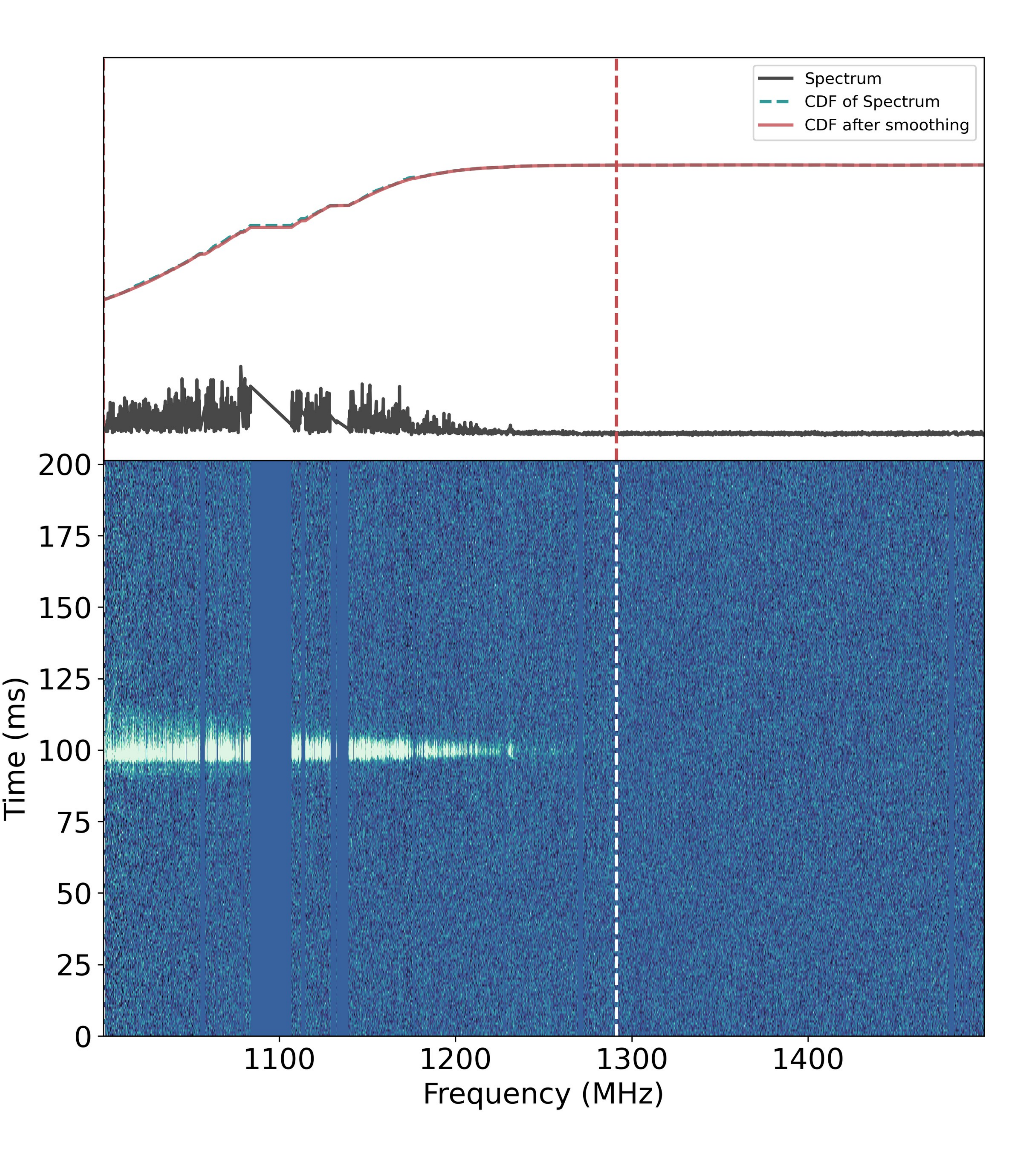}
  \includegraphics[width=0.49\textwidth]{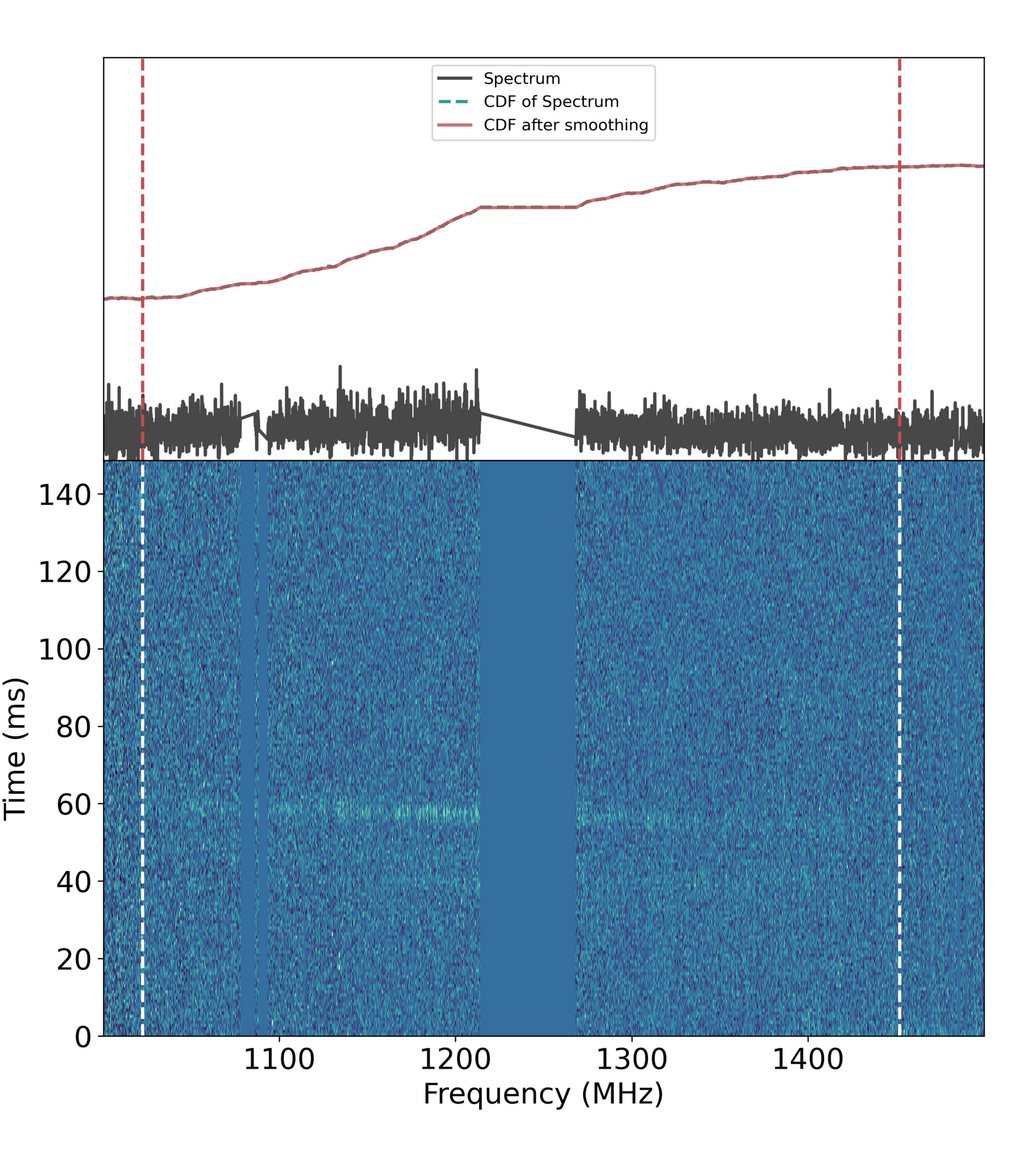}
  \caption{Two examples of calculated frequency burst ranges, both originating from FRB~20220912A \citep{2023ApJ...955..142Z}. In the upper panel, the black solid line represents the spectrum of the burst, the green dashed line denotes the CDF of the spectrum, and the red solid line indicates the smoothed CDF after filtering.}
  \label{fig:cdf_freq}
\end{figure*}

\begin{figure*}[!htp]
  \centering
  \includegraphics[width=0.49\textwidth]{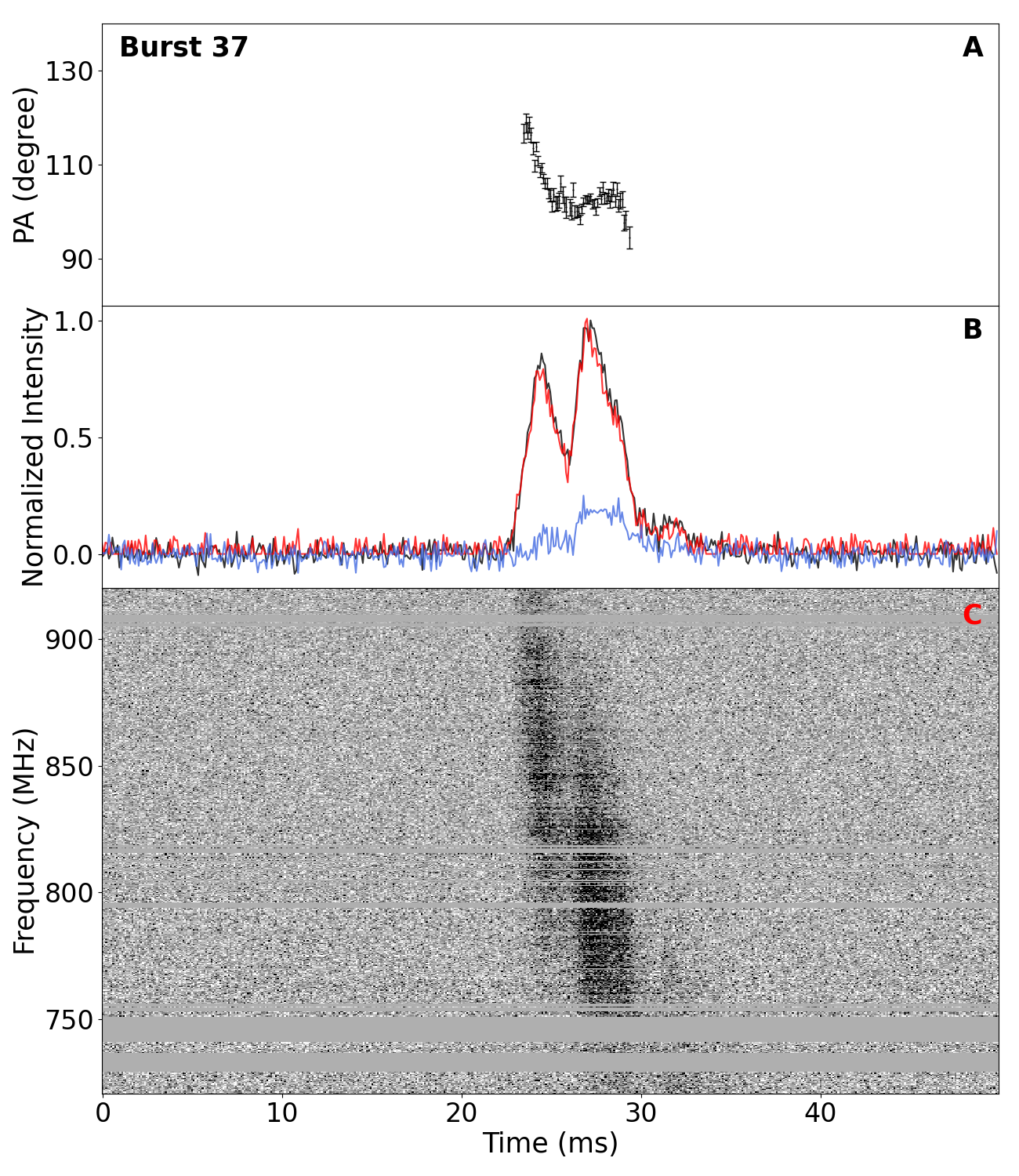}
  \includegraphics[width=0.49\textwidth]{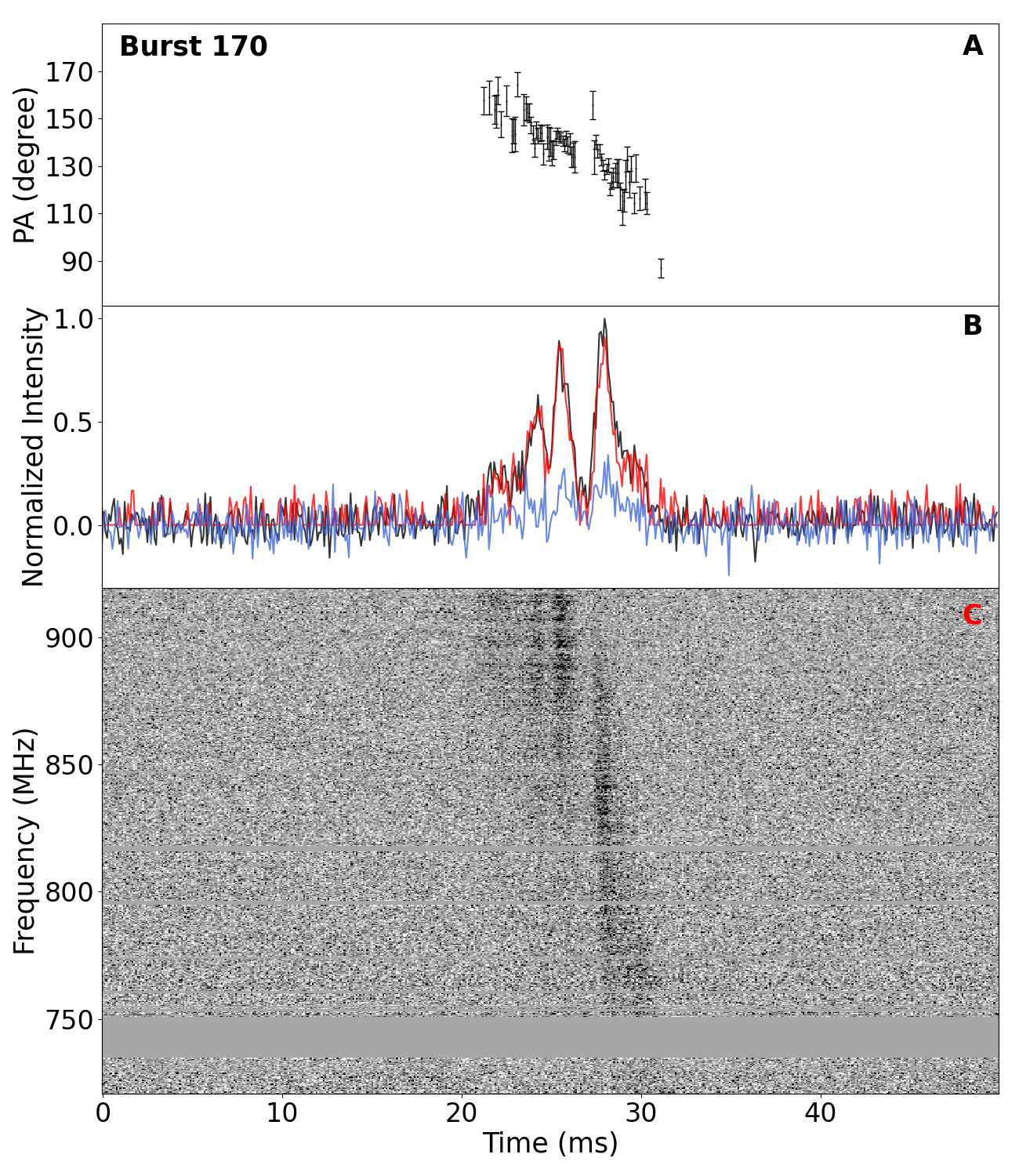}\\
  \includegraphics[width=0.49\textwidth]{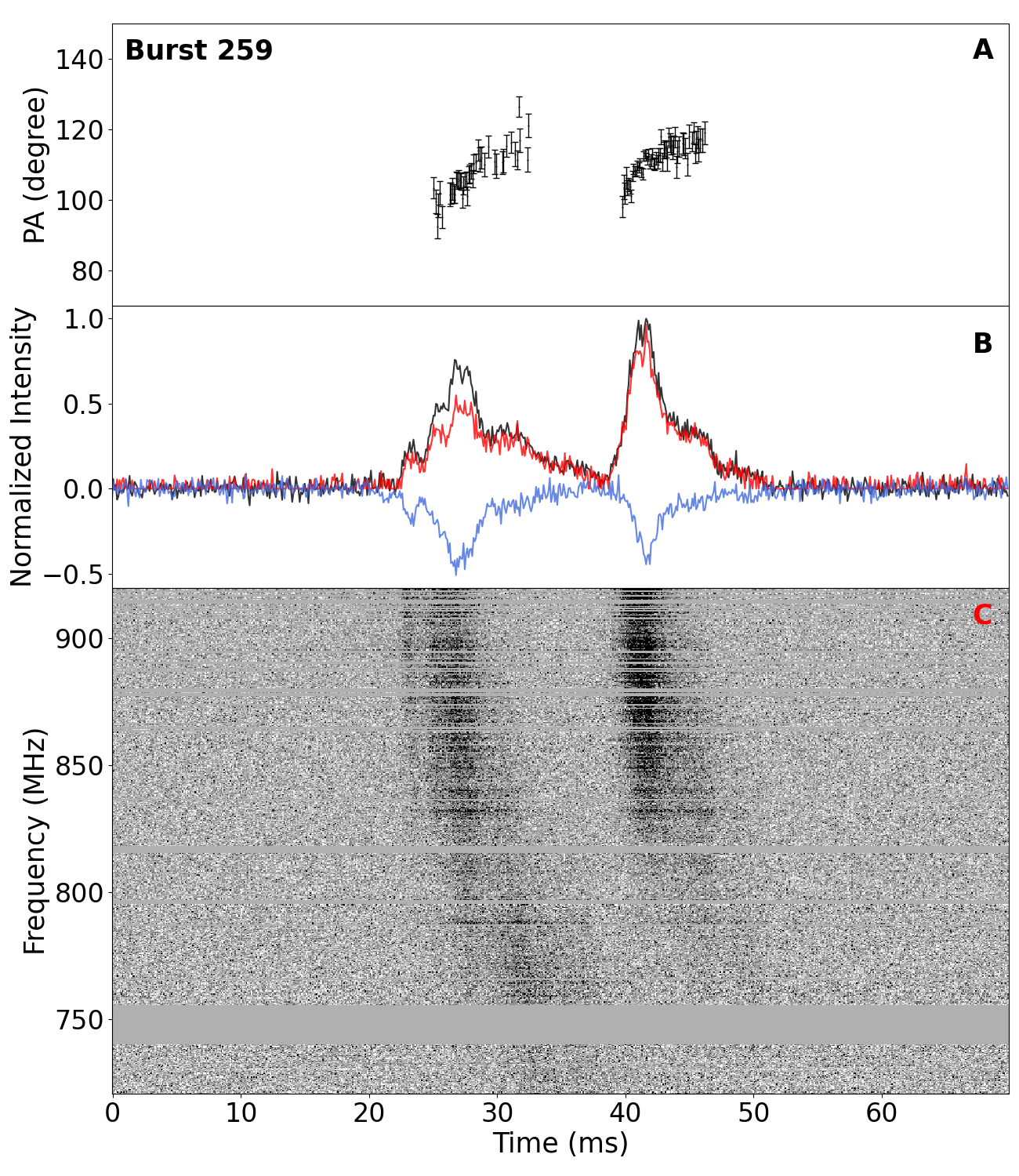}
  \includegraphics[width=0.49\textwidth]{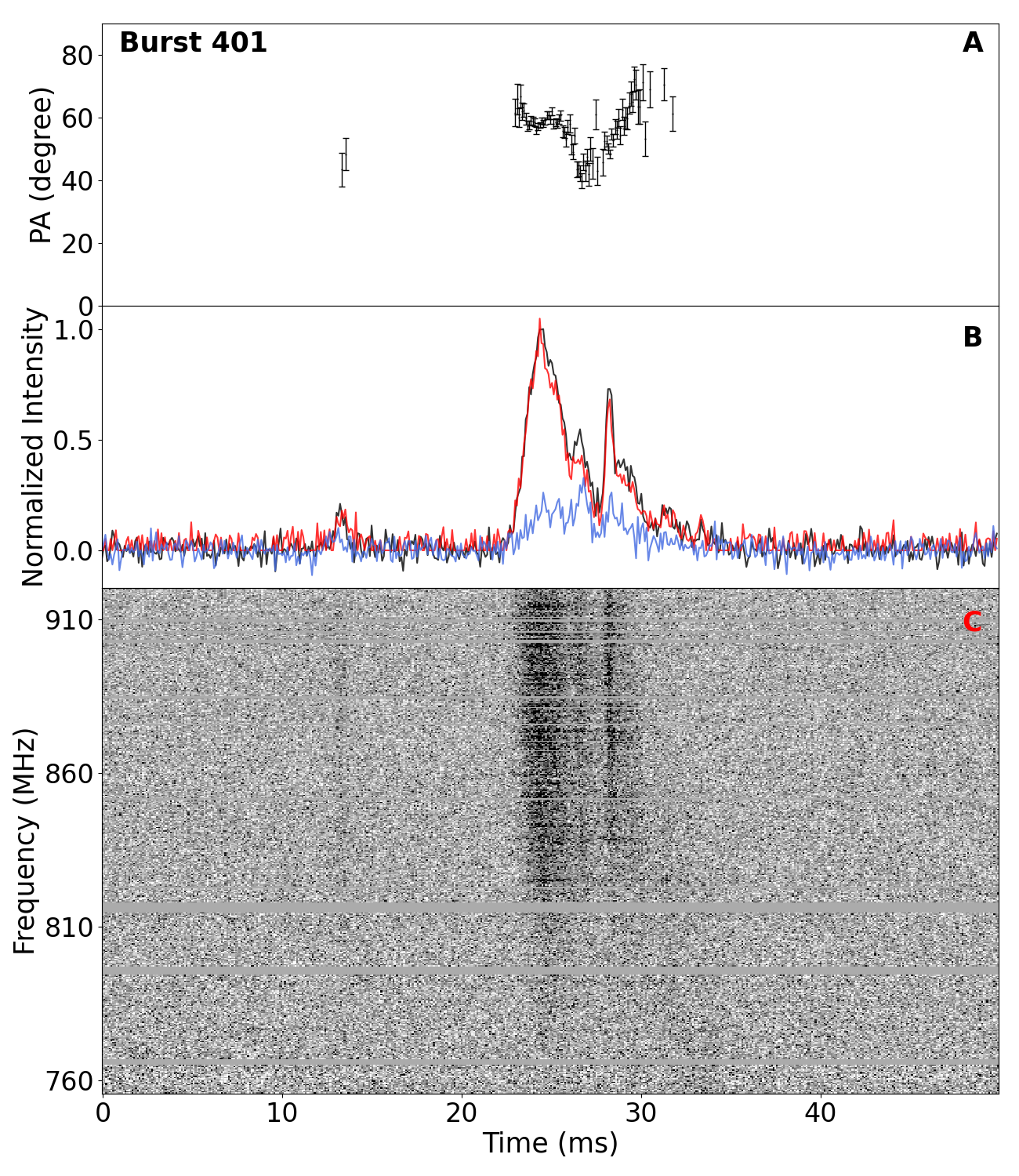}
  \caption{Polarization position angle and intensity profiles, along with dynamic spectra, for four bursts from FRB~20240114A. In each panel: sub-panel A displays the polarization position angles; sub-panel B presents the burst profiles, with lines representing total intensity (black, normalized to a peak value of 1.0), linear polarization (red), and circular polarization (blue); sub-panel C shows the dynamic spectra.}
  \label{fig:spec}
\end{figure*}

The calibration process was implemented through the {\sc pac} module of the PSRCHIVE software \citep{2004PASA...21..302H}, utilizing differential on- and off-source observations of the primary calibrator 3C 394 to quantify the absolute flux density of the P-band receiver's integrated noise diode. Prior to the main observations, the noise diode was observed again at the position of FRB~20240114A, and these data were used to flux calibrate each detected burst. Following standard procedures for coherent search mode calibration, the resulting flux densities were multiplied by a factor of 20 to account for an online processing scaling factor.

Radio frequency interference (RFI) was mitigated using the PSRCHIVE software package \citep{2004PASA...21..302H}.
A median filter was applied to each burst in the frequency domain using the paz command, and RFI was manually mitigated in each burst using the pazi command. For DM optimization, we used {\sc DM-power}\footnote{https://github.com/hsiuhsil/DM-power} \citep{2022arXiv220813677L}.
Polarization calibration was performed by correcting for the gain and phase differences between the receptors. This was achieved through separate measurements of a noise diode signal aligned at a $45^{\circ}$ angle relative to the linear receptors, modeled with the single-axis approach in the PSRCHIVE software package. The rotation measure (RM) was determined using the Stokes QU-fitting method \citep{2012MNRAS.421.3300O}, focusing on regions where the signal-to-noise ratio exceeded 3.
Due to challenges in calculating RM from weaker signals or bursts with narrow emission bandwidths, RM values were successfully measured for 301 of the detected bursts. We corrected the linear polarization by derotating it using the measured RM.

The degrees of linear and circular polarization were calculated for bursts with detected RM. The presence of noise leads to an overestimation of the measured linear polarization. To mitigate this, we employ the frequency-averaged, de-biased total linear polarization \citep{2001ApJ...553..341E,2020MNRAS.497.3335D}\footnote{\citet{2020MNRAS.497.3335D} corrected a typographical error in \cite{2001ApJ...553..341E}.}:
\begin{equation} \label{eq:L_de-bias}
    L_{{\mathrm{de\mbox{-}bias}}} =
    \begin{cases}
      \sigma_I \sqrt{\left(\frac{L_{i}}{\sigma_I}\right)^2 - 1} & \text{if $\frac{L_{i}}{\sigma_I} > 1.57$} \\
      0 & \text{otherwise} ,
    \end{cases}
\end{equation}
where $\sigma_I$ is the off-pulse standard deviation of Stokes $I$, and $L_i$ represents the frequency-averaged linear polarization of time sample $i$. The degree of linear polarization is defined as $L/I$, where $L=\Sigma_{i} L_{\mathrm{de\mbox{-}bias},i}$ and $I=\Sigma_{i}I_i$, with the summation performed over all time samples within a burst. Similarly, the degree of circular polarization is defined as $V/I$, where $V=\Sigma_{i} V_i$ and $V_i$ is the frequency-averaged circular polarization of time sample $i$. Uncertainties in the linear and circular polarization fractions are quantified using the following equations:
\begin{equation} \label{eq:uncertainty}
    \sigma_{\rho/I} = \frac{\sqrt{N+N\frac{\rho^2}{I^2}}}{I}\sigma_{I},
\end{equation}
where $N$ denotes the number of time samples within the burst (with signal exceeding the noise by 3$\sigma$), and $\rho$ represents $L$ or $V$ for the linear and circular polarization fractions, respectively.

While the composite spectra of FRBs exhibit a distinct power-law distribution, the power-law model proves insufficient in characterizing the narrowband spectral features of individual bursts. In previous studies, the determination of burst frequency ranges has relied on Gaussian model fitting, as seen in the analyses of FRB~20121102A\citep{2021ApJ...922..115A}, FRB~20201124A\citep{2022RAA....22l4001Z}, and FRB~20220912A\citep{2023ApJ...955..142Z}. However, the Gaussian fitting method struggles to reliably determine the frequency ranges of bursts with complex spectral structures or low signal-to-noise ratios. \cite{2024A&A...683A.183M} proposed a method for determining the dominant frequency range by applying Kadane’s algorithm to spectral data to identify the subarray with the highest energy. This technique significantly enhances sensitivity to weak emissions through optimized localization of high-energy spectral components, demonstrating particular efficacy in low signal-to-noise regimes characteristic of distant or faint astrophysical transients. In this study, we introduce a novel method to calculate the lower and upper frequency boundaries of bursts within the observed frequency band, aiming to mitigate the impact of noise on the accuracy of frequency range calculations for weak pulses.

The boundaries of the burst frequency range are systematically adjusted based on the behavior of the first derivative of the cumulative distribution function (CDF) of the spectrum. The spectrum is calculated as the cumulative sum of the pulse range for each channel, represented as:
\begin{equation} \label{eq:spec}
    P(f)_{i} = \sum_{t_{\mathrm{min}}}^{t_{\mathrm{max}}} s_i(t),
\end{equation}
where $s_i(t)$ represents the flux density at the $t$-th phase bin within the frequency $i$-th channel. The pulse range is determined either by cross-correlating a Gaussian template with the burst pulse profile or through interactive definition. Following this, the cumulative distribution function $\mathrm{CDF}(\nu)$ is derived from the power spectrum as:
\begin{equation} \label{eq:spec_cdf}
\mathrm{CDF}(\nu) = \sum_{\nu_{\mathrm{min}}}^{\nu} P(f)_{i},~~~~ \nu_{\mathrm min} \leq \nu \leq \nu_{\mathrm{max}}
\end{equation}
where $\nu_{\mathrm{min}}$ is the minimum observation frequency. Subsequently, the CDF data is normalized using smoothing filters in conjunction with the Asymmetrically Reweighted Penalized Least Squares algorithm \citep[arPlS]{2015Ana...140..250B}. The first derivative of the smoothed and normalized spectral CDF is computed to identify the frequency indices that define the start and end points of the frequency range, ensuring it encompasses the region where the CDF equals 0.5. If the first derivative remains consistently positive below this region, the lower boundary is set to the minimum frequency in the data; conversely, if it is positive above this region, the upper boundary is established at the maximum frequency. Finally, uncertainties in the lower and upper frequency limits are estimated using the bootstrap method \citep{bootstrap_err}. We tested the algorithm using burst data from FRB 20220912A observed with FAST \citep{2023ApJ...955..142Z}, with two examples illustrated in Figure \ref{fig:cdf_freq}. The results indicate that this algorithm accurately calculates the frequency range of the bursts, and removing interference bands within the bursts does not affect the computed results.

The time of arrivals, DM, RM, degrees of linear, circular polarization and lower frequency and upper boundaries of bursts are listed in Table \ref{tab:burst}.

\section{Results and Discussion} \label{sec:results}

In this study, we detected a total of 437 bursts from FRB~20240114A during two GBT observation sessions.
The average burst rate is approximately 161 $\rm hr^{-1}$. During the 1.38-hour observation on March 1, 2024, 359 bursts from FRB~20240114A were detected, corresponding to a burst rate of 260 bursts per hour, which is the highest burst rate recorded to date with the GBT or any other (sub)-100-meter radio telescope.
The burst information obtained from the observational analysis can be downloaded from \href{https://blinkverse.alkaidos.cn/publication/download}{the Blinkverse website}\footnote{\url{https://blinkverse.alkaidos.cn/publication/download}}.

\begin{figure}[htbp]
\centering
\includegraphics[width=0.98\linewidth]{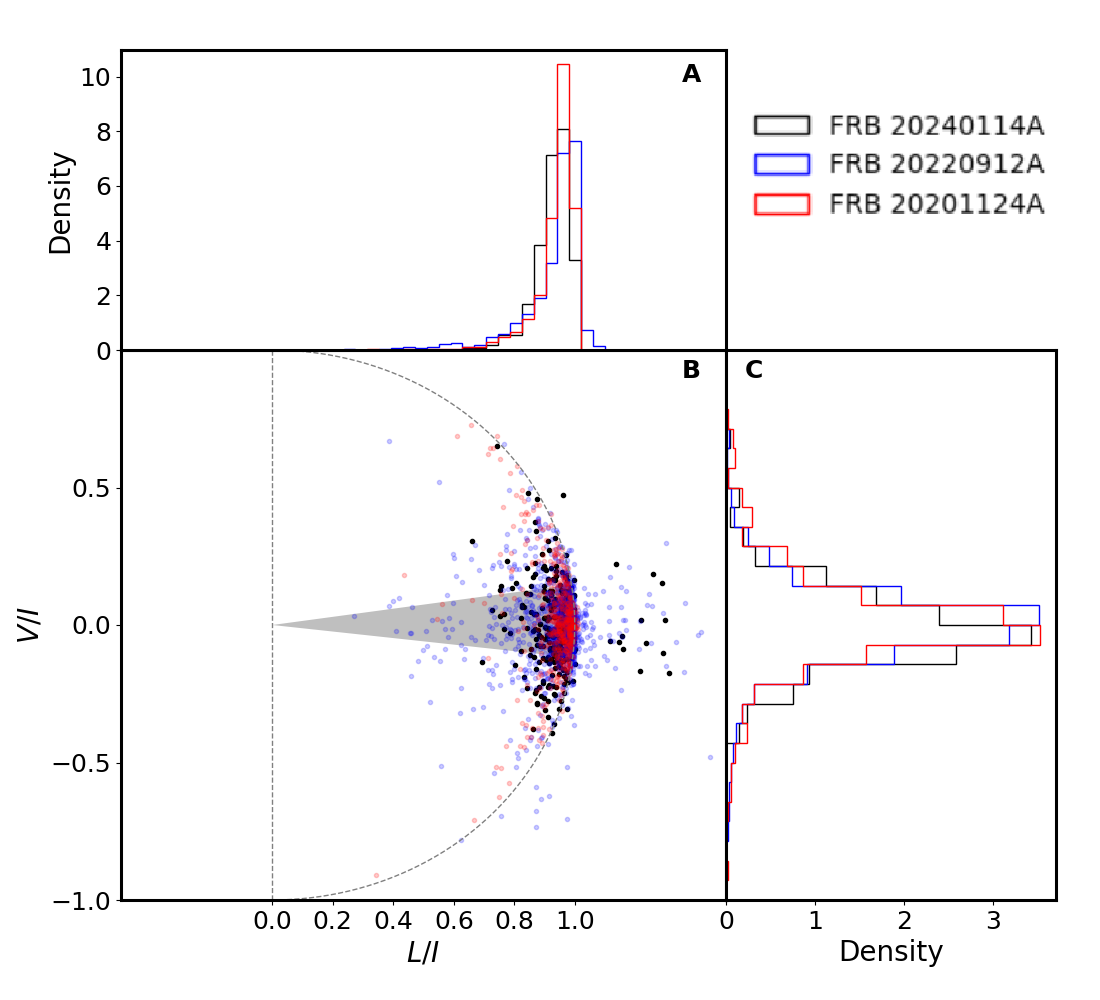}
\caption{\textbf{A} Histograms of linear polarization fraction $L/I$ for FRB~20240114A, FRB~20220912A, and FRB~20201124A. \textbf{B} Correlations between linear polarization fraction $L/I$ and circular polarization fraction $V/I$ for FRB~20240114A, FRB~20220912A, and FRB~20201124A. The grey dashed semicircles denote the areas in which the degree of polarization $P/I=\sqrt{L^2+V^2}/I\leq1$. The grey sectors cover 68\% of bursts from FRB~20240114A. \textbf{C} Histograms of circular polarization fraction $V/I$ for FRB~20240114A, FRB~20220912A, and FRB~20201124A.}
\label{fig:L_V}
\end{figure}

Considering the contribution of the ionosphere to the RMs, the values were corrected using a Python code, \texttt{IonFR}, along with global ionospheric map (GIM) products \citep{2013ascl.soft03022S}. These GIM products are available in the IONosphere EXchange (IONEX) format on NASA's website\footnote{cddis.nasa.gov/archive/gnss/products/ionex/}. 
For our GBT observations, the RM contribution from the ionosphere was $2.4 \pm 0.4$ rad~m$^{-2}$ on February 23 and $2.9 \pm 0.4$ rad~m$^{-2}$ on March 1, respectively. After correcting for the ionospheric contribution to the RM and excluding bursts with RM error bars larger than 5\,rad\,m$^{-2}$, the average RM for the February 23 observation was \(347.0 \pm 1.0~ \mathrm{rad~m^{-2}}\), while the average RM for the March 1 observation was \(353.7 \pm 0.6 \, \mathrm{rad~m^{-2}}\). 

In the 301 RM-detected bursts, approximately 81\% show a linear polarization fraction greater than 90\%.
This means that FRB~20240114A is not depolarized much due to multipath
propagation. Depolarization due to multipath
propagation can be described by \cite{2022Sci...375.1266F}:
\begin{equation}
\label{eq:rmscatter}
f_{\rm{RM~scattering}} \equiv 1 - \exp{(-2\lambda^4\sigma^2_{\mathrm{RM}})} \>\>,
\end{equation}
where $f_{\rm{RM~scattering}}$ represents the fractional reduction in linear polarization amplitude, $\sigma_{\mathrm{RM}}$ is the standard deviation of the RM, and $\lambda$ is the wavelength. Considering the narrow bandwidth of our observation and the high linear polarization fraction, we provide an upper limit on the $\sigma_{\mathrm{RM}}$ rather than measuring $\sigma_{\mathrm{RM}}$ precisely. We note that burst 139 has a linear polarization fraction of $97\pm2\%$, observed between 721\,MHz and 919\,MHz. Using Eq. \ref{eq:rmscatter}, we place an upper limit on $\sigma_{\mathrm{RM}}$ of $0.9 \pm 0.3 \, \mathrm{rad \, m^{-2}}$.    

Of the 301 bursts with detected RMs, 14\% exhibit circular polarization with a signal-to-noise ratio greater than 5. The highest observed absolute fractional circular polarization is $65.2\pm 3.7$\%. FRB~20240114A is the fifth repeating FRB observed to exhibit circular polarization. The previously reported repeating FRBs with circular polarization are all active sources: FRB~20201124A \citep{2022Natur.609..685X}, FRB~20190520B, FRB~20121102A \citep{2022SciBu..67.2398F}, and FRB~20220912A \citep{zhang2023, feng2024}.

Polarization angle swings were observed in some bursts of FRB~20240114A, similar to other repeating FRBs, including FRB~20180301A \citep{luo2020}, FRB~20201124A \citep{2022Natur.609..685X, Niu_2024}, and FRB~20220912A \citep{feng2024}.
Figure~\ref{fig:spec} presents the polarization angles, burst profiles, and dynamic spectra of representative bursts. Notably, bursts 36 and 394 clearly display polarization angle swings.  Full polarization position angle, multi-component intensity profiles, and frequency-time dynamic spectra for all 437 detected bursts are compiled in Appendix Figure \ref{fig:appendix1}.

Figure \ref{fig:L_V} shows the distributions of linear polarization fraction ($L/I$) and circular polarization fraction ($V/I$) for FRB~20240114A. Fig. \ref{fig:L_V}\textbf{A} presents the histograms of $L/I$, while Fig. \ref{fig:L_V}\textbf{B} illustrates the correlation between $L/I$ and $V/I$, with the grey dashed semicircles indicating $P/I = \sqrt{L^2 + V^2}/I \leq 1$ and grey sectors covering 68\% of the bursts from FRB~20240114A. Fig. \ref{fig:L_V}\textbf{C} shows the histograms of $V/I$. Additionally, we show the distributions of linear and circular polarization fractions for the sample from FRB~20220912A in \cite{zhang2023} and the sample from FRB~20201124A in \cite{jiangraa}. It appears that the three repeating FRBs have similar distributions of linear and circular polarization fractions. This suggests that the radiation mechanisms of the three sources are identical.

\begin{figure}[htbp]
\centering
\includegraphics[width=0.98\linewidth]{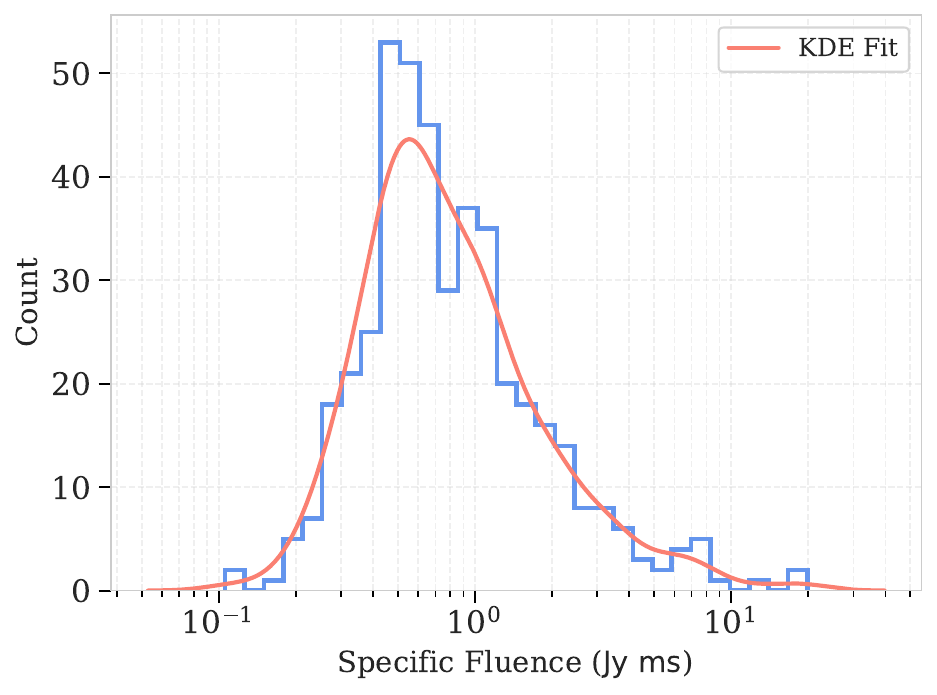}
\caption{Histograms of specific fluence for FRB~20240114A. The red line represents the kernel density estimation (KDE) of the specific fluence distribution.} 
\label{fig:fluence_hist}
\end{figure}

We calculated the specific fluence by integrating the pulse signal over the determined frequency range of each burst. This method ensures that the fluence is derived directly from the observed emission within the identified frequency band, making it particularly relevant for narrowband FRBs, where fluence measurements are sensitive to the chosen frequency integration limits. The specific fluence distribution was then analyzed using a log-transformed kernel density estimation (KDE) approach, providing deeper insights into its statistical properties, as shown in Figure~\ref{fig:fluence_hist}. The Kolmogorov–Smirnov (K-S) test yielded a p-value of 0.4939, suggesting a reasonable fit to the observed data. The mean and median fluence values, derived from KDE sampling, are 0.7923 $\mathrm{Jy~ms}$ and 0.7105 $\mathrm{Jy~ms}$, respectively, indicating a right-skewed distribution. This asymmetry suggests the presence of high-fluence bursts that shift the mean upwards. The standard deviation of 2.2974 $\mathrm{Jy~ms}$ further confirms the broad dispersion in fluence values. 
The right-skewed nature of the distribution is consistent with previously studied repeating FRBs \citep{2021Natur.598..267L,2022RAA....22l4001Z,2023ApJ...955..142Z,2024ApJ...974..296F}, where bursts exhibit a wide range of fluences.

\begin{figure}[htbp]
\centering
\includegraphics[width=0.98\linewidth]{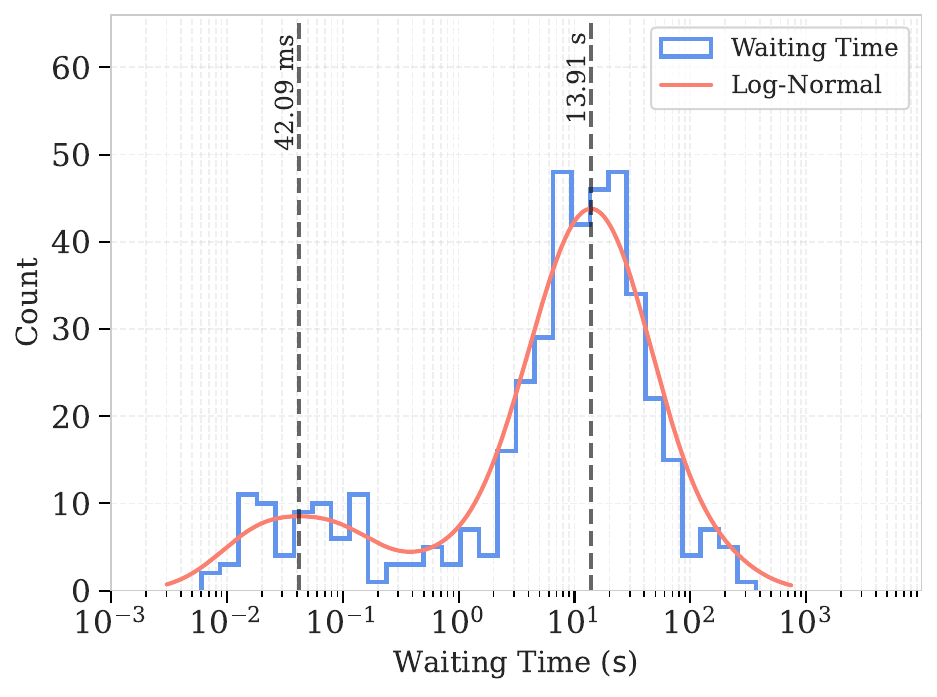}
\caption{Waiting time distribution of FRB~20240114A. The blue step plot represents the distribution of waiting times, while the red curve denotes the fit using two log-normal components.} 
\label{fig:waiting_time_hist}
\end{figure}

We analyzed the waiting-time distribution of FRB20240114A and found that it exhibits a bimodal structure, similar to previously studied repeating FRBs such as FRB20201124A and FRB20220912A. We modeled the distribution using two log-normal functions, yielding peaks at approximately 42.09 ms and 13.91 s, as shown in Figure \ref{fig:waiting_time_hist}. The K-S test for the log-normal fit gives a p-value of 0.23, indicating that while the model captures the overall trend, some deviations remain. The right peak represents the activity level of the FRB source over the statistical period, while the left peak at 42.09 ms is comparable to the characteristic short waiting times observed in other repeating FRBs. 
Notably, FRB~20220912A exhibits peaks at 51 ms and 18 s \citep{2023ApJ...955..142Z}, while FRB~20201124A shows a short waiting-time peak at 39 ms in \cite{2022Natur.609..685X} and 51 ms in \cite{2022RAA....22l4002Z}, along with a longer peak at 10 s. The presence of short waiting times on the order of tens of milliseconds in multiple sources suggests a potentially intrinsic timescale associated with the burst emission mechanism, while the longer waiting-time peaks may reflect variations in the activity cycles of different FRBs sources.

\section{Conclusions}\label{sec:conclusion}

We report on the hyperactivity of FRB~20240114A and its polarization characteristics. Our key findings are as follows:

\begin{itemize}
\setlength{\itemsep}{3pt}
\item[1.] A total of 437 bursts were detected in total with the GBT at 720 to 920\,MHz. During the 1.38-hour observation on March 1, 2024, 365 bursts were detected, corresponding to a burst rate of 264 events per hour. This is the highest burst rate observed to date with the GBT or any other (sub)-100-meter radio telescope.
\item[2.] The mean RM values were \(347.0 \pm 1.0 \, \mathrm{rad~m^{-2}}\) on February 23 and \(353.7 \pm 0.6 \, \mathrm{rad~m^{-2}}\) on March 1. The combined mean RM was \(350.3 \pm 0.2 \, \mathrm{rad~m^{-2}}\).
\item[3.] Out of the 301 bursts with RM detection, 81\% bursts exhibit a linear polarization fraction greater than 90\%, and 14\% of the bursts have circular polarization with signal-to-noise ratio $>$ 5. We placed an upper limit on $\sigma_{\mathrm{RM}}$ of $0.9 \pm 0.3 \, \mathrm{rad \, m^{-2}}$. Some bursts from FRB~20240114A show downward drift in frequency and polarization angle swings. 
\item[4.] We have compared the linear polarization fraction of FRB~20240114A with those of the repeating FRB~20201124A and FRB~20220912A. We find that all three sources show similar distributions in their linear and circular polarization fractions. Our findings imply that the radiation mechanisms of these three sources are consistent with each other.
\item[5.] We propose a novel approach for determining the frequency range of bursts based on their intrinsic spectral properties. This method operates independently of specific spectral models and remains robust even in the presence of excised channels affected by radio frequency interference, ensuring reliable performance across diverse observational conditions.
\item[5.] We analyzed the fluence and waiting-time distributions of FRB~20240114. The analysis revealed that the fluence distribution exhibits a right-skewed characteristic. The waiting-time distribution exhibits a bimodal structure with peaks at 42.09 ms and 13.91 s, similar to other repeating FRBs. These results suggest intrinsic emission timescales and variability in activity cycles. The observed bimodal features in both fluence and waiting times indicate the potential presence of multiple burst populations or emission mechanisms, highlighting the need for further studies to constrain the physical origins of repeating FRBs.

\end{itemize}

\section*{Acknowledgments}
\begin{acknowledgments}
We thank Jinchen Jiang for providing the data in Jiang et al. 2022. This work is supported by National Key R\&D Program of China No. 2023YFE0110500, National Natural Science Foundation of China grant No.\ 11988101, 12203045, 11725313, 12041303, by the Leading Innovation and Entrepreneurship Team of Zhejiang Province of China grant No. 2023R01008, by Key R\&D Program of Zhejiang grant No. 2024SSYS0012, by the CAS Youth Interdisciplinary Team, the Youth Innovation Promotion Association CAS (id. 2021055), and by the Cultivation Project for FAST Scientific Payoff and Research Achievement of CAMS-CAS. Di Li is a New Cornerstone investigator. The Green Bank Observatory is a facility of the National Science Foundation operated under cooperative agreement by Associated Universities, Inc.
\end{acknowledgments}

\bibliography{frb20240114}{}
\bibliographystyle{aasjournal}


\appendix
\setlength{\tabcolsep}{3.5pt} 


\renewcommand{\thefigure}{A\arabic{figure}}
\setcounter{figure}{0}
\begin{figure*}
    \centering
    \includegraphics[width=.88\textwidth]{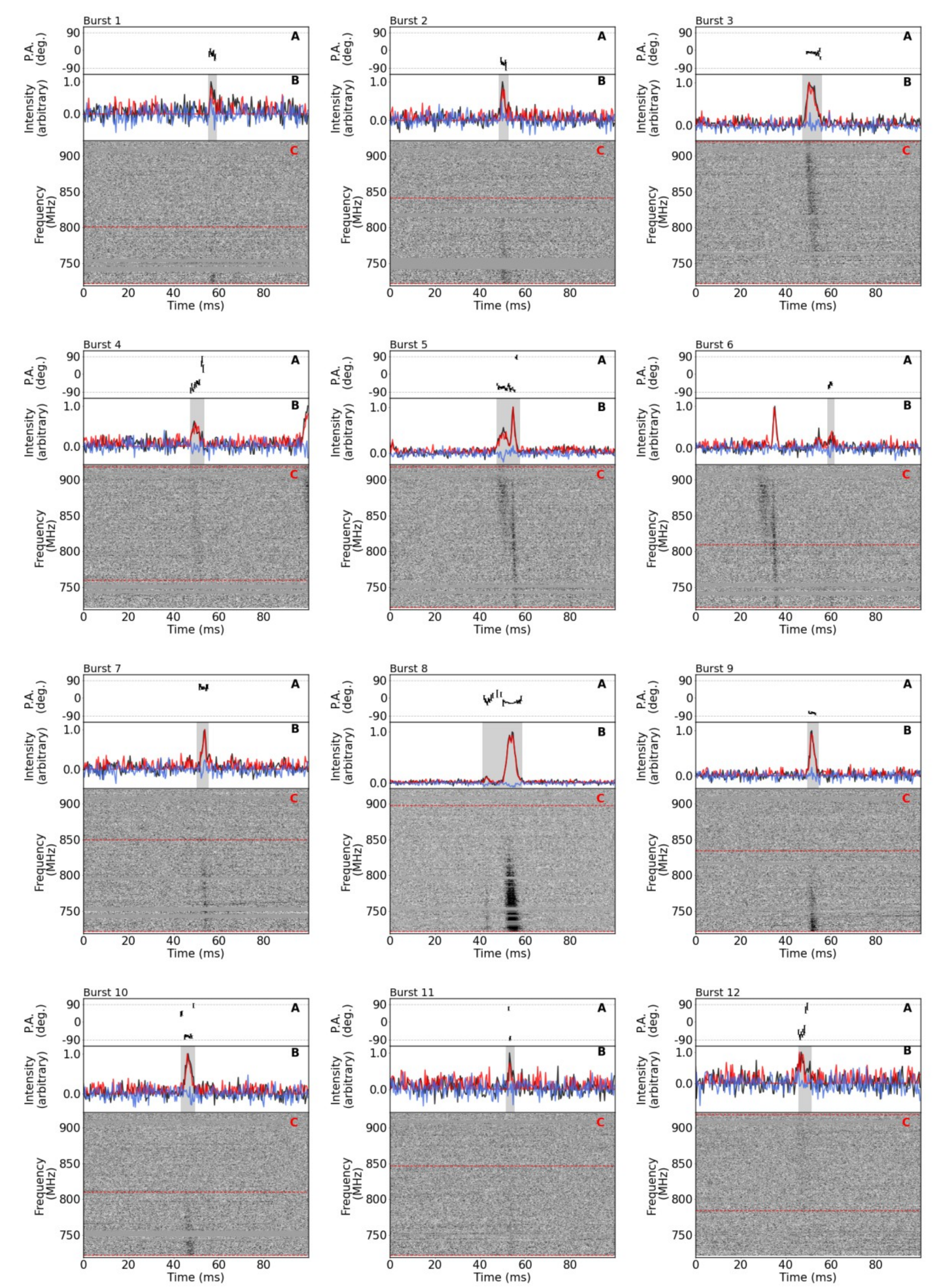}
    \caption{Polarization position angle and intensity profiles of FRB~20240114A with dynamic spectra, presented chronologically. Sub-panel A shows the polarization position angles (omitted for bursts without significant linear polarization measurements at $2-\sigma$ level).; sub-panel B shows the polarization burst profiles with lines indicating total intensity (black, normalized to a peak value of 1.0), linear polarization (red) and circular polarization (blue), the temporal extent of the burst is represented by gray shaded areas; sub-panel C shows the 2 dimensional dynamic spectrum overfrequency and time, where red dashed lines demarcate spectral boundaries.}
    \label{fig:appendix1}
\end{figure*}

\renewcommand{\thefigure}{A\arabic{figure} (Continued.)}
\addtocounter{figure}{-1}
\begin{figure*}
\centering
\includegraphics[width=.95\textwidth]{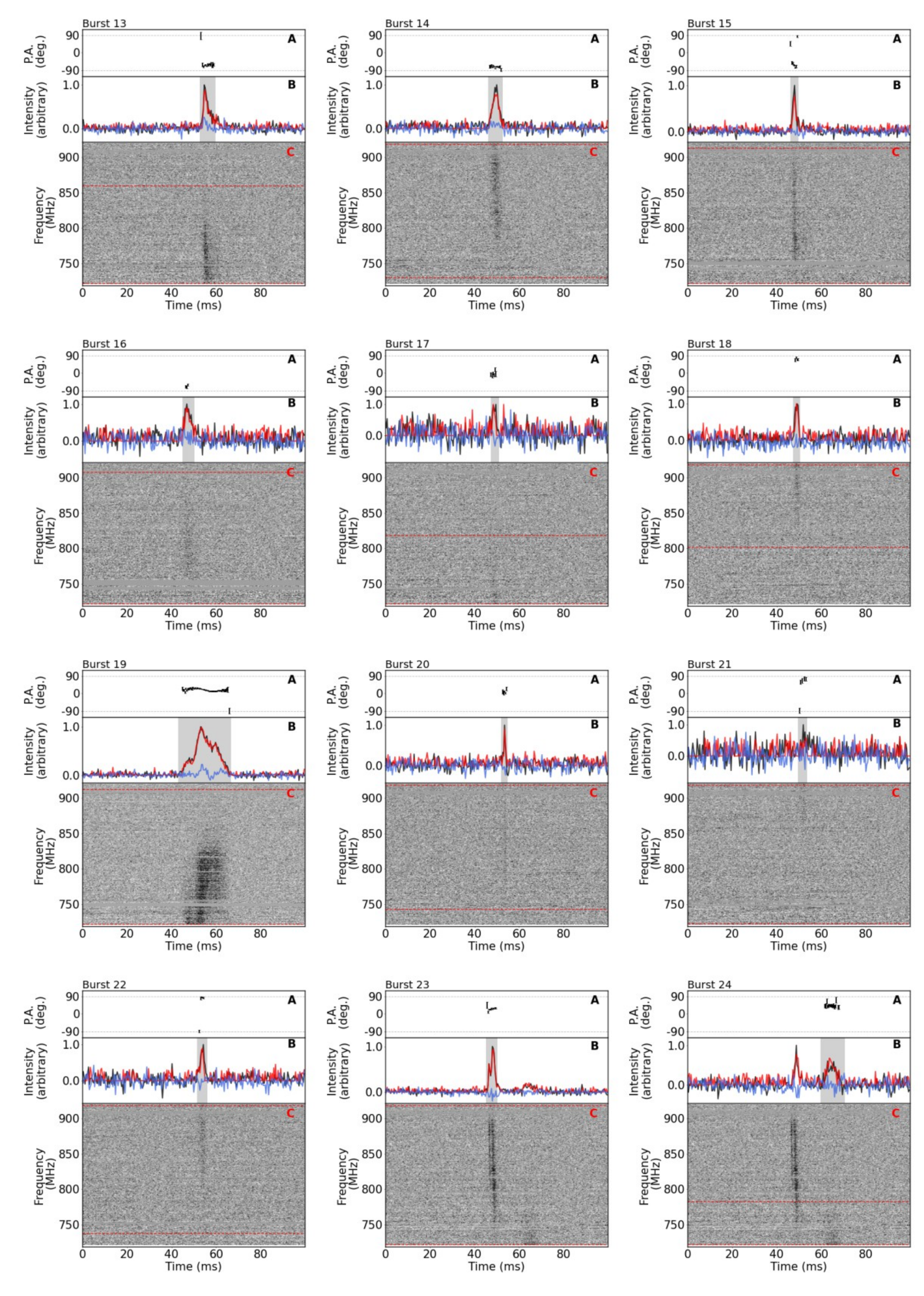}
\caption{}
\end{figure*}

\renewcommand{\thefigure}{A\arabic{figure} (Continued.)}
\addtocounter{figure}{-1}
\begin{figure*}
\centering
\includegraphics[width=.95\textwidth]{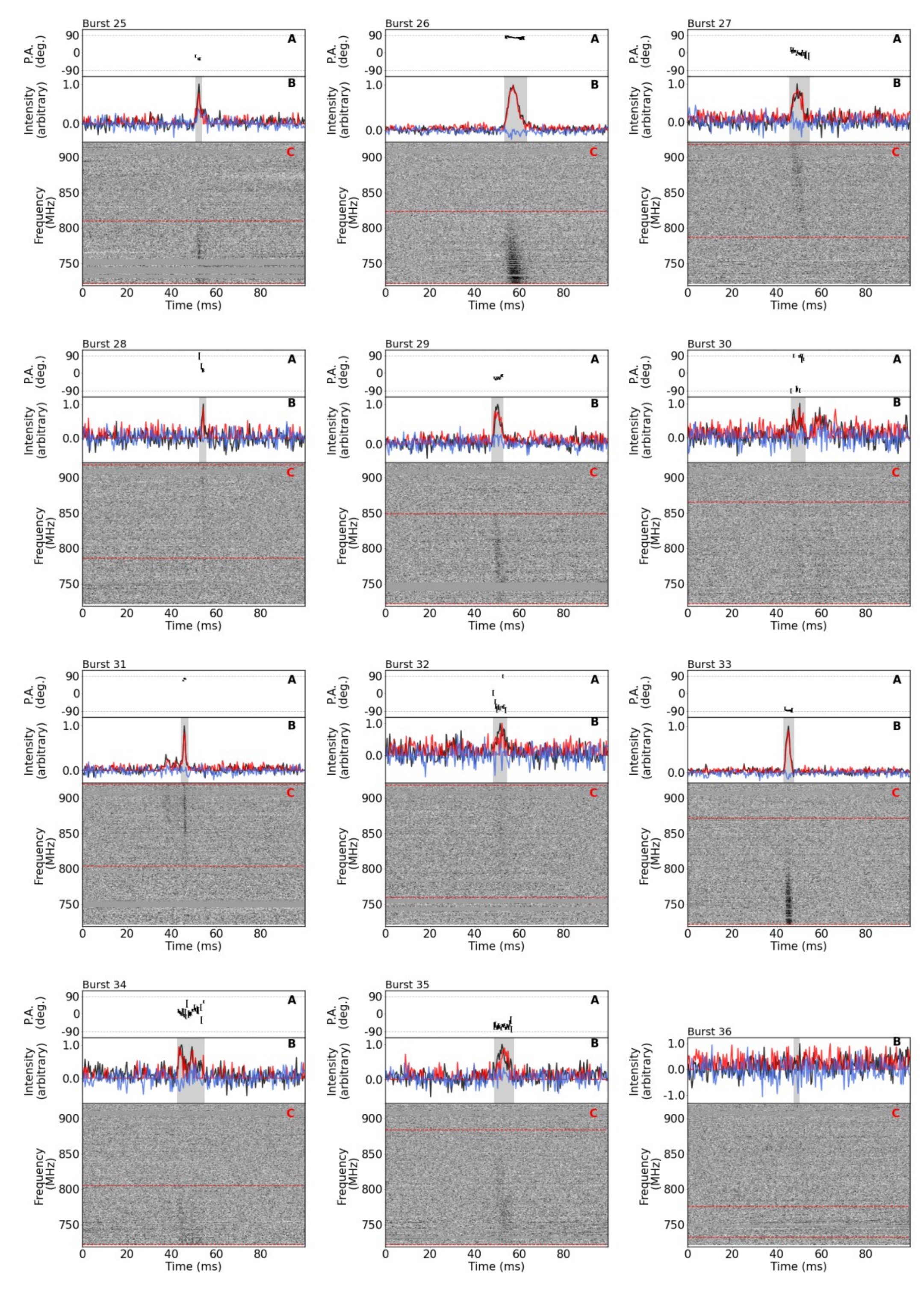}
\caption{}
\end{figure*}

\renewcommand{\thefigure}{A\arabic{figure} (Continued.)}
\addtocounter{figure}{-1}
\begin{figure*}
\centering
\includegraphics[width=.95\textwidth]{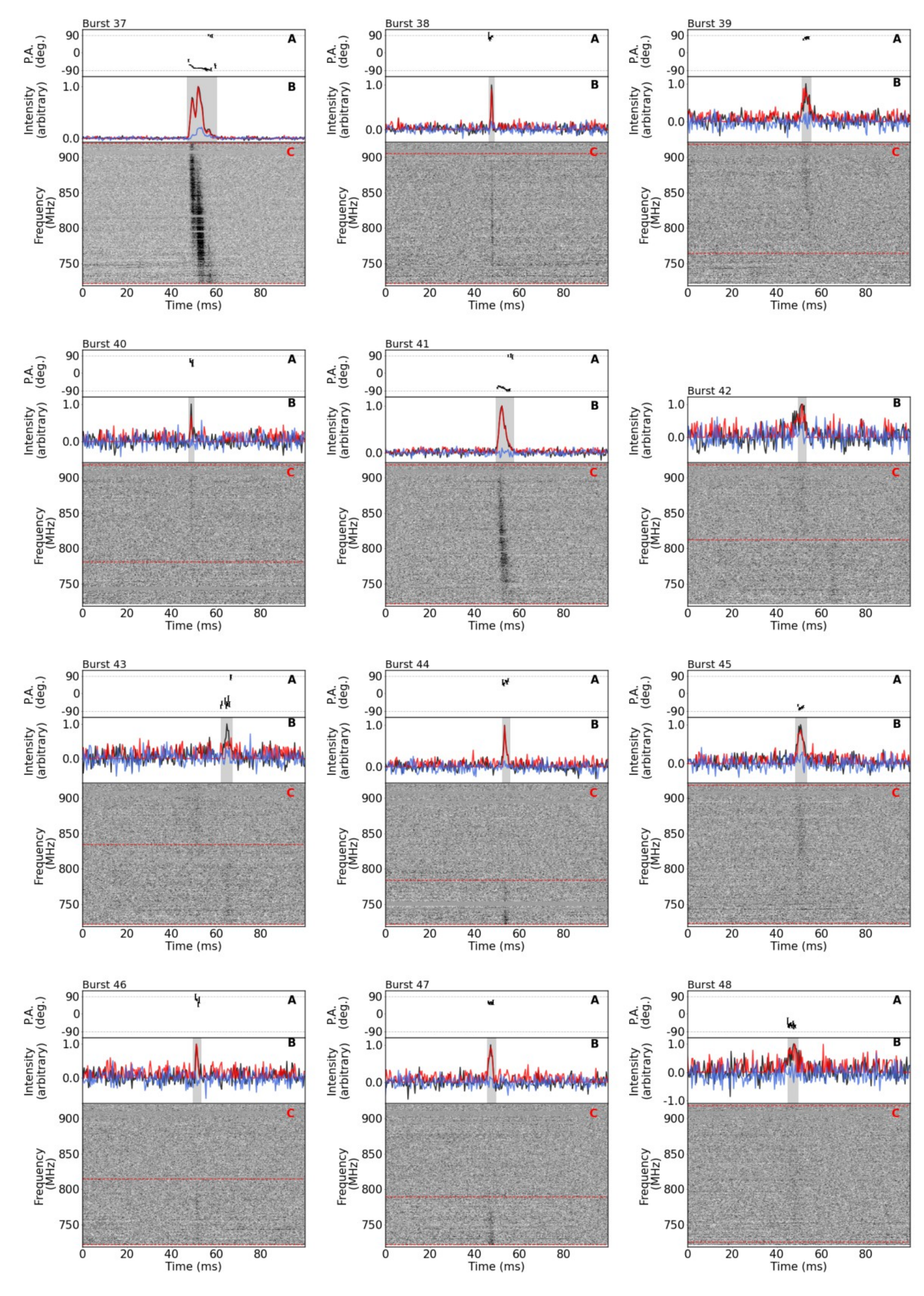}
\caption{}
\end{figure*}

\renewcommand{\thefigure}{A\arabic{figure} (Continued.)}
\addtocounter{figure}{-1}
\begin{figure*}
\centering
\includegraphics[width=.95\textwidth]{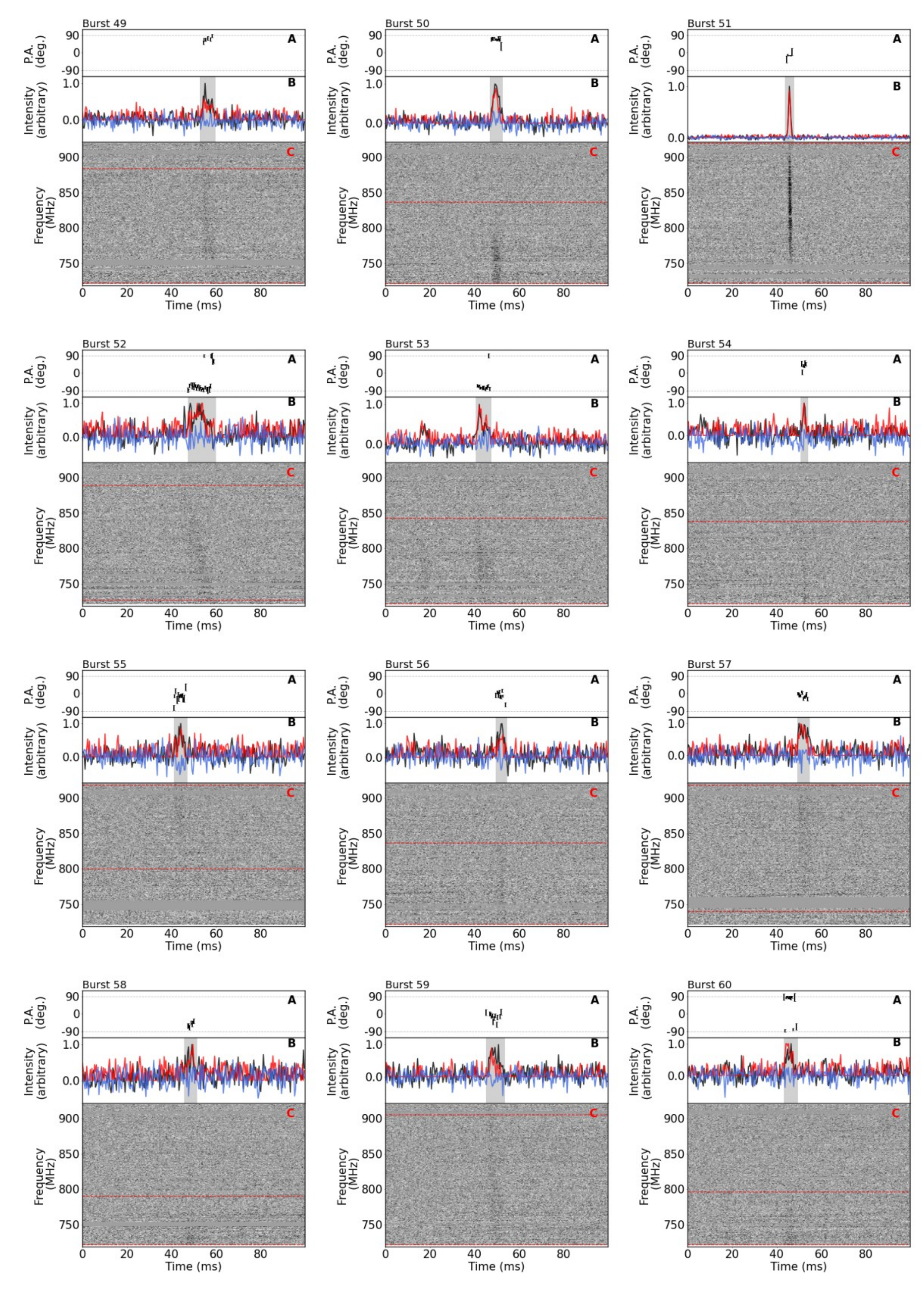}
\caption{}
\end{figure*}

\renewcommand{\thefigure}{A\arabic{figure} (Continued.)}
\addtocounter{figure}{-1}
\begin{figure*}
\centering
\includegraphics[width=.95\textwidth]{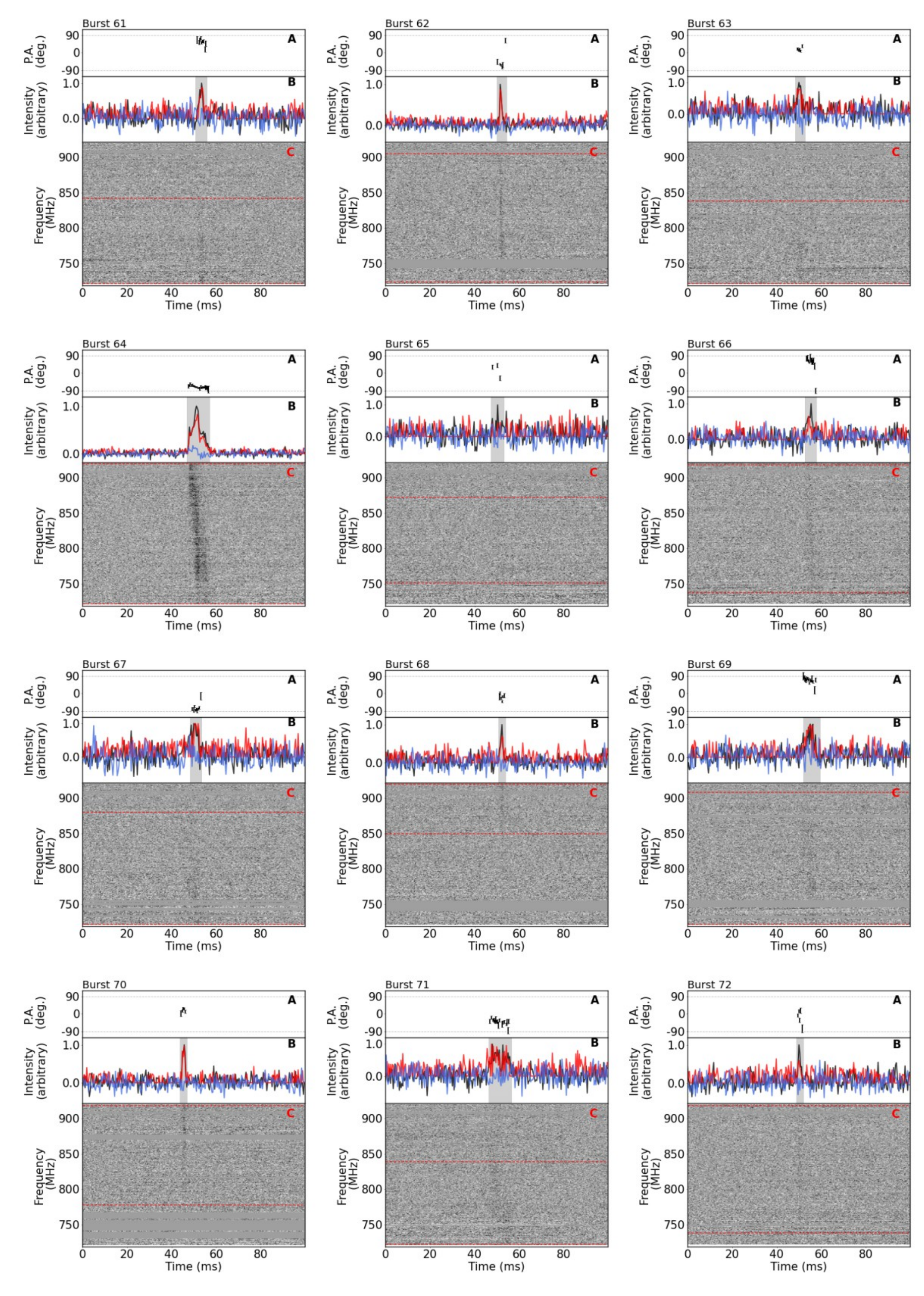}
\caption{}
\end{figure*}

\renewcommand{\thefigure}{A\arabic{figure} (Continued.)}
\addtocounter{figure}{-1}
\begin{figure*}
\centering
\includegraphics[width=.95\textwidth]{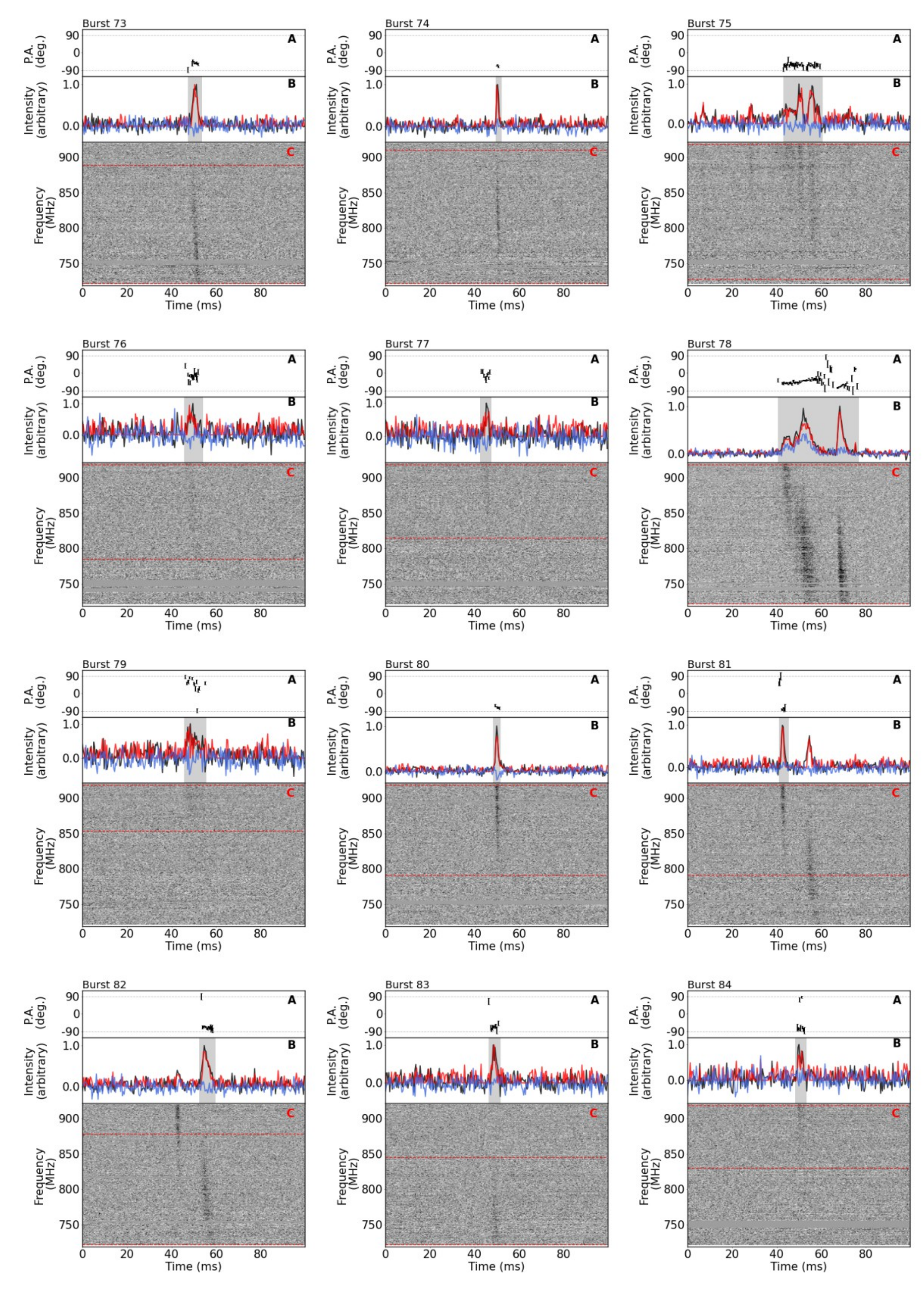}
\caption{}
\end{figure*}

\renewcommand{\thefigure}{A\arabic{figure} (Continued.)}
\addtocounter{figure}{-1}
\begin{figure*}
\centering
\includegraphics[width=.95\textwidth]{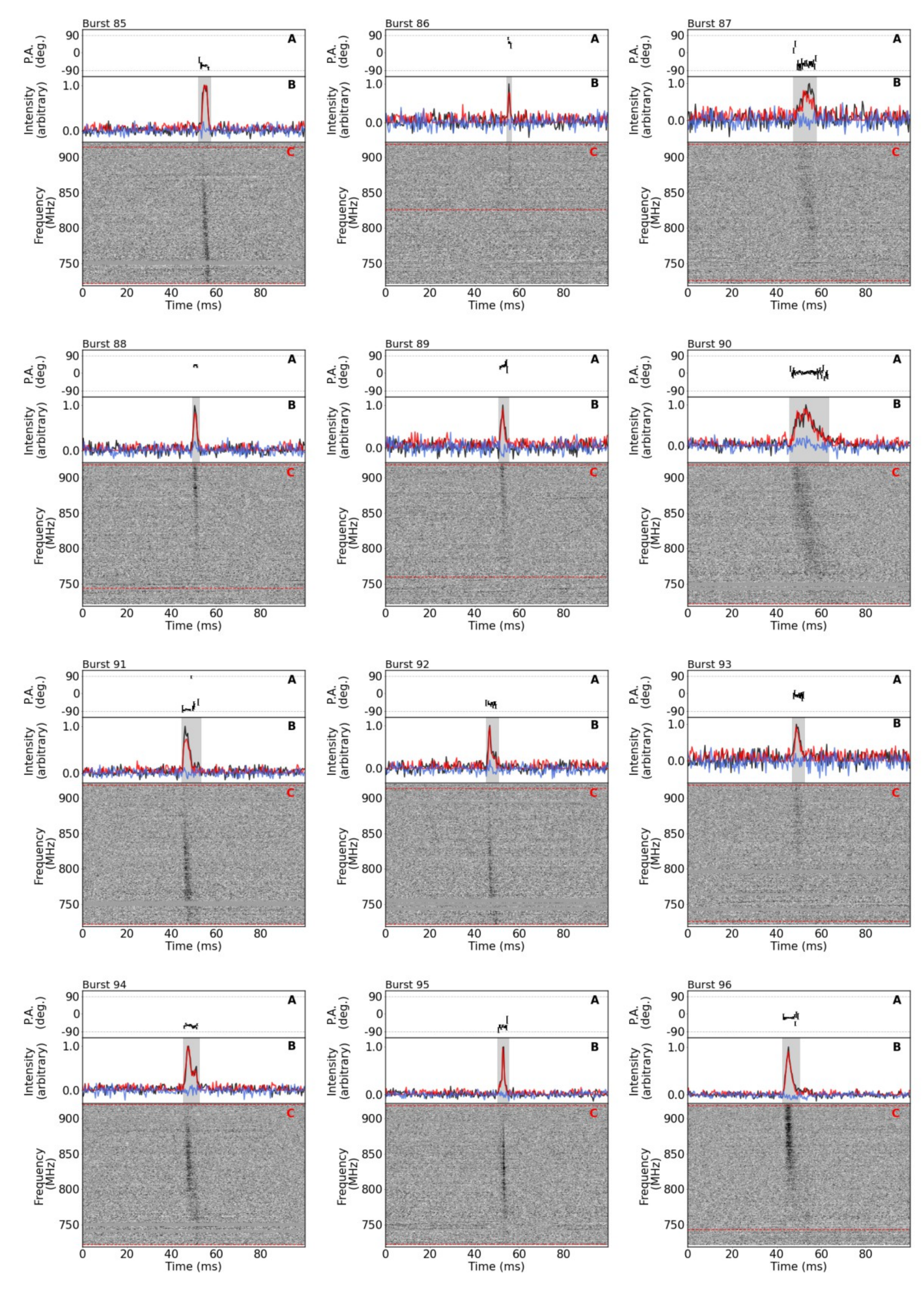}
\caption{}
\end{figure*}

\renewcommand{\thefigure}{A\arabic{figure} (Continued.)}
\addtocounter{figure}{-1}
\begin{figure*}
\centering
\includegraphics[width=.95\textwidth]{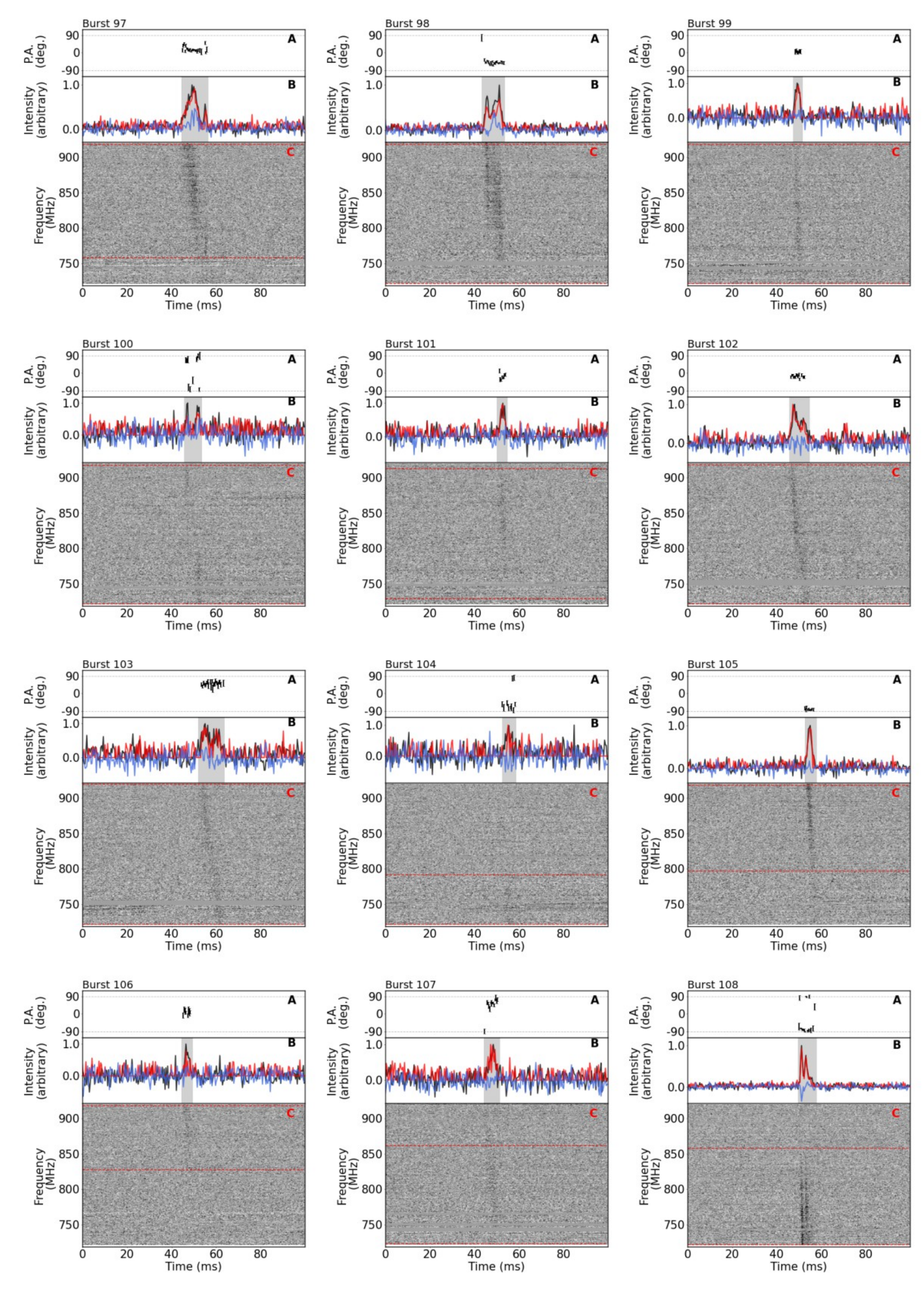}
\caption{}
\end{figure*}

\renewcommand{\thefigure}{A\arabic{figure} (Continued.)}
\addtocounter{figure}{-1}
\begin{figure*}
\centering
\includegraphics[width=.95\textwidth]{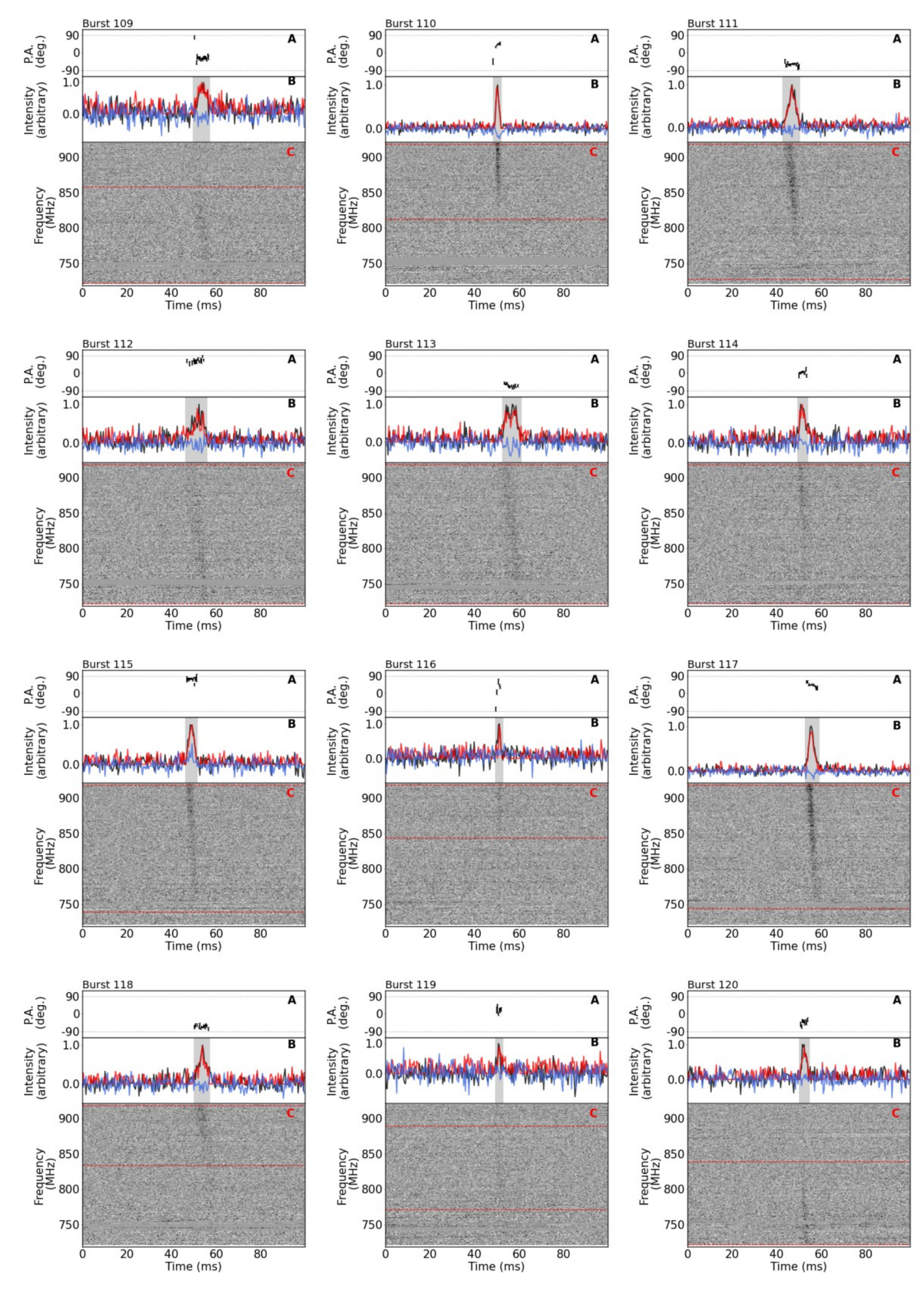}
\caption{}
\end{figure*}

\renewcommand{\thefigure}{A\arabic{figure} (Continued.)}
\addtocounter{figure}{-1}
\begin{figure*}
\centering
\includegraphics[width=.95\textwidth]{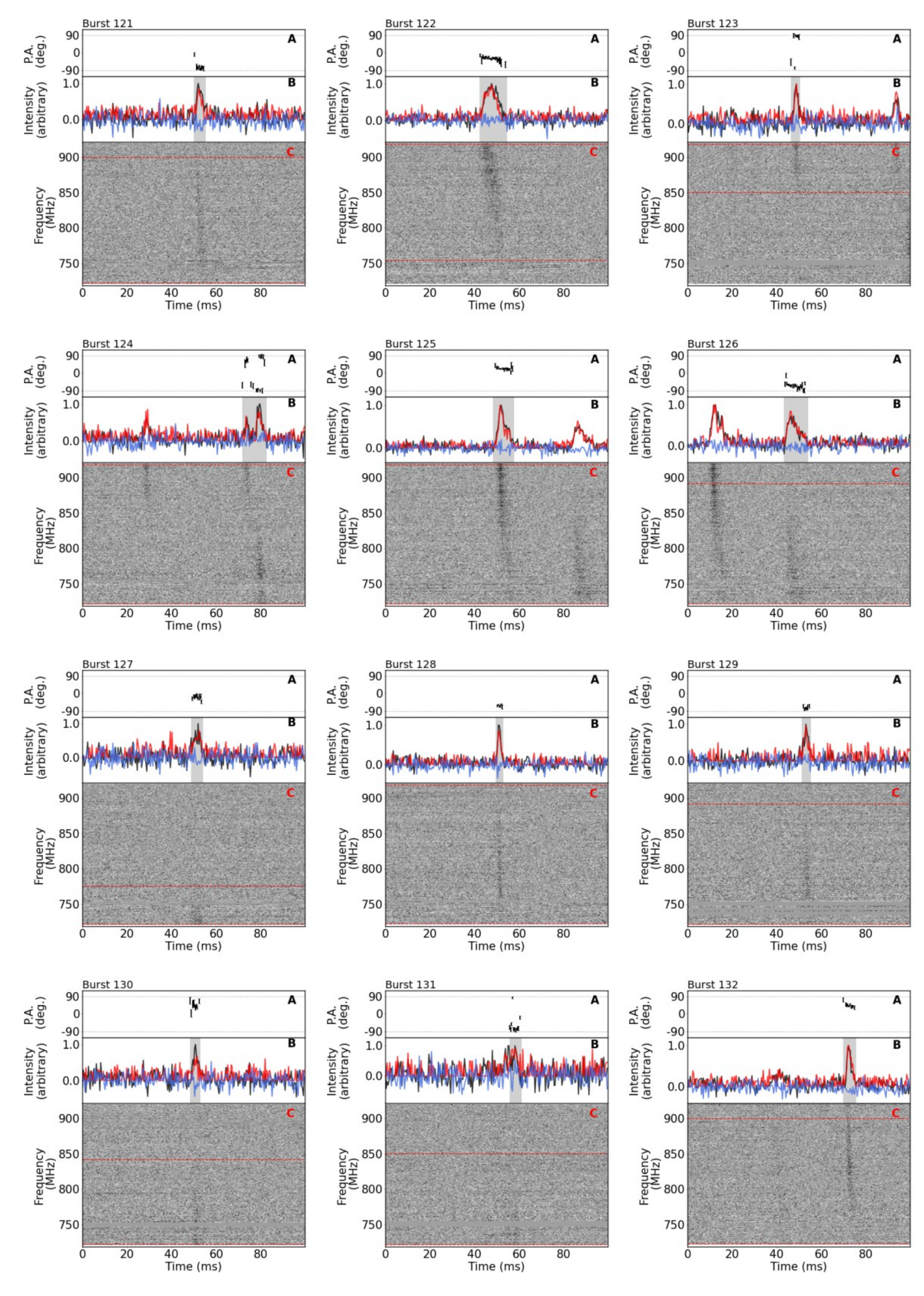}
\caption{}
\end{figure*}

\renewcommand{\thefigure}{A\arabic{figure} (Continued.)}
\addtocounter{figure}{-1}
\begin{figure*}
\centering
\includegraphics[width=.95\textwidth]{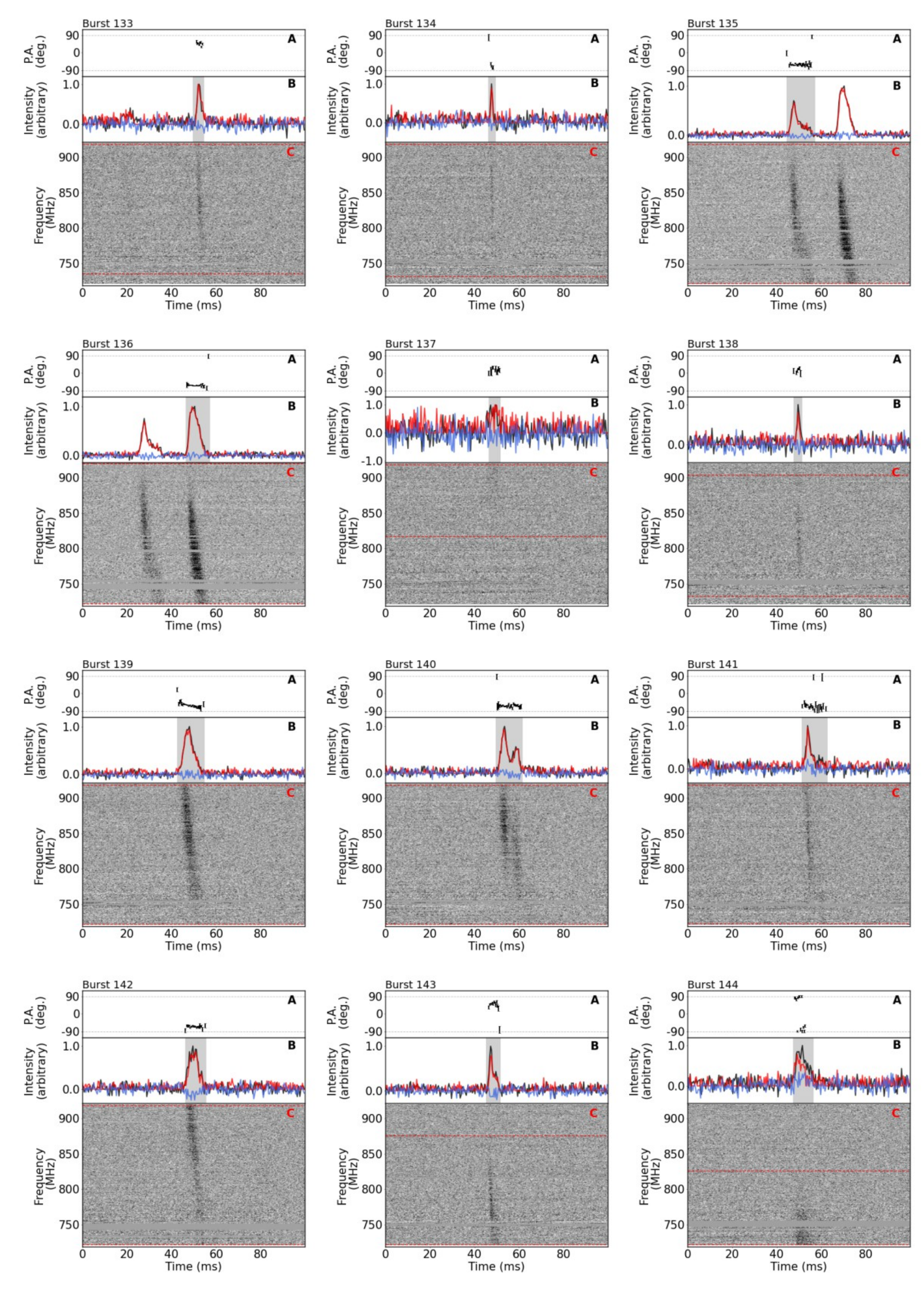}
\caption{}
\end{figure*}

\renewcommand{\thefigure}{A\arabic{figure} (Continued.)}
\addtocounter{figure}{-1}
\begin{figure*}
\centering
\includegraphics[width=.95\textwidth]{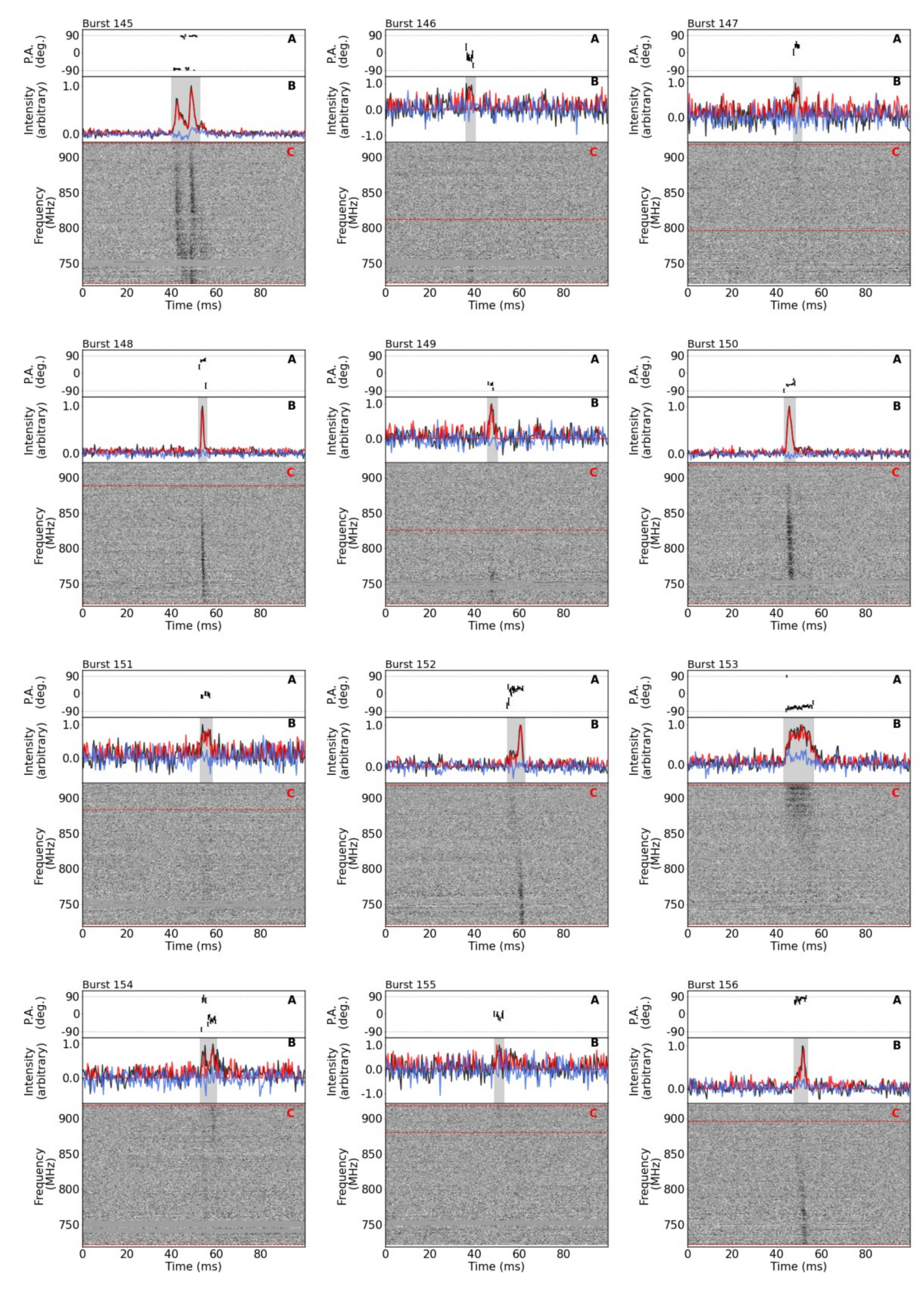}
\caption{}
\end{figure*}

\renewcommand{\thefigure}{A\arabic{figure} (Continued.)}
\addtocounter{figure}{-1}
\begin{figure*}
\centering
\includegraphics[width=.95\textwidth]{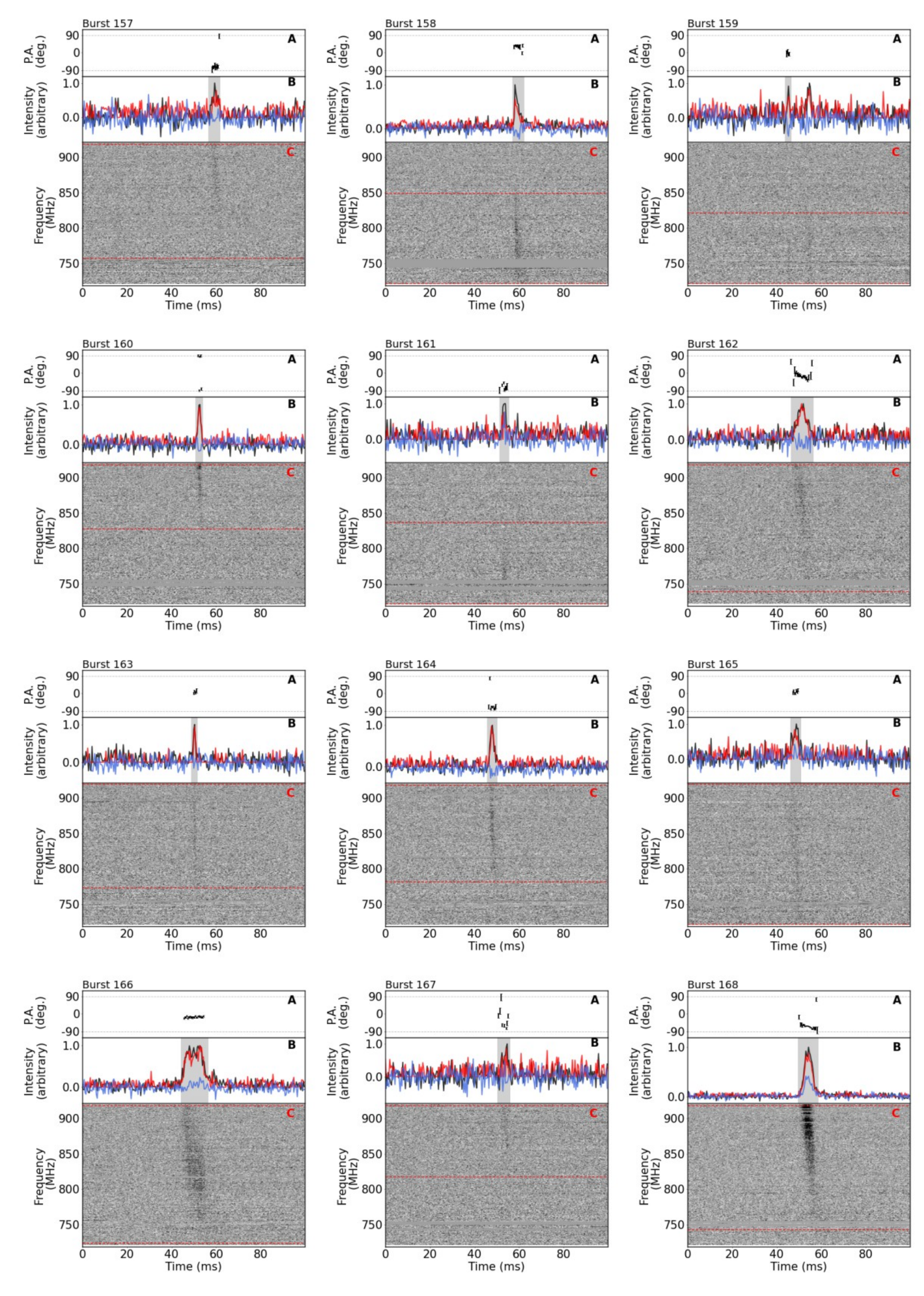}
\caption{}
\end{figure*}

\renewcommand{\thefigure}{A\arabic{figure} (Continued.)}
\addtocounter{figure}{-1}
\begin{figure*}
\centering
\includegraphics[width=.95\textwidth]{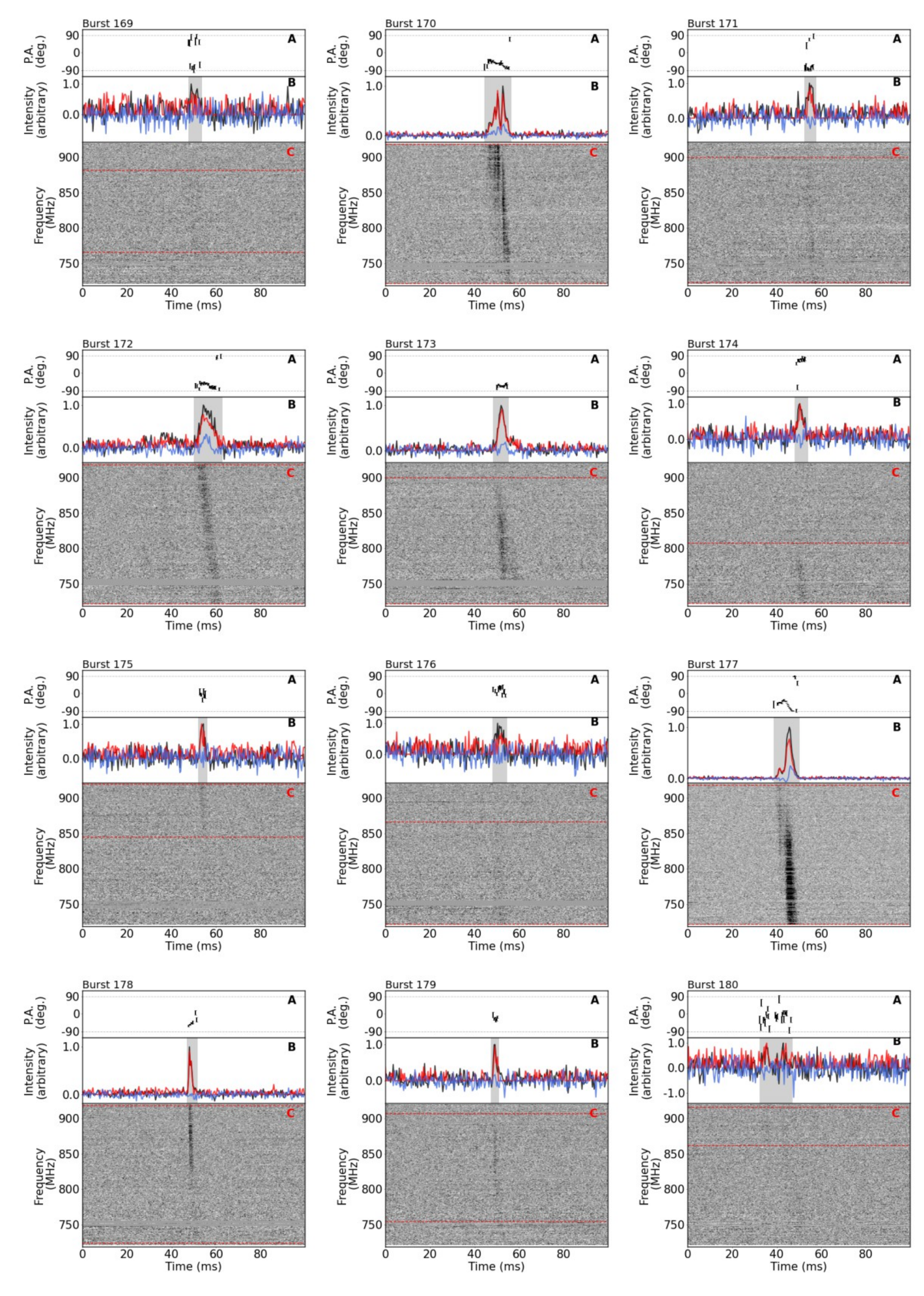}
\caption{}
\end{figure*}

\renewcommand{\thefigure}{A\arabic{figure} (Continued.)}
\addtocounter{figure}{-1}
\begin{figure*}
\centering
\includegraphics[width=.95\textwidth]{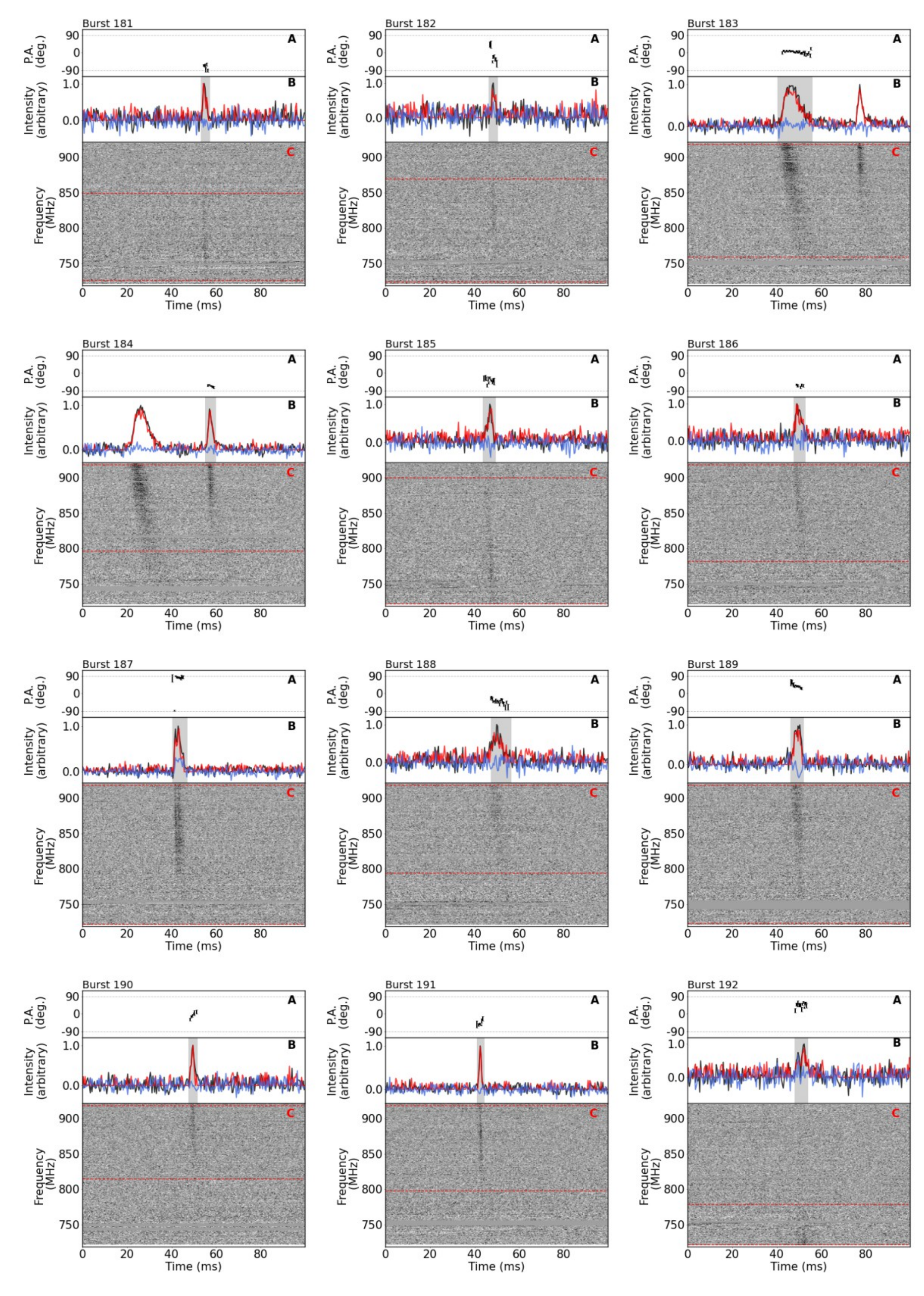}
\caption{}
\end{figure*}

\renewcommand{\thefigure}{A\arabic{figure} (Continued.)}
\addtocounter{figure}{-1}
\begin{figure*}
\centering
\includegraphics[width=.95\textwidth]{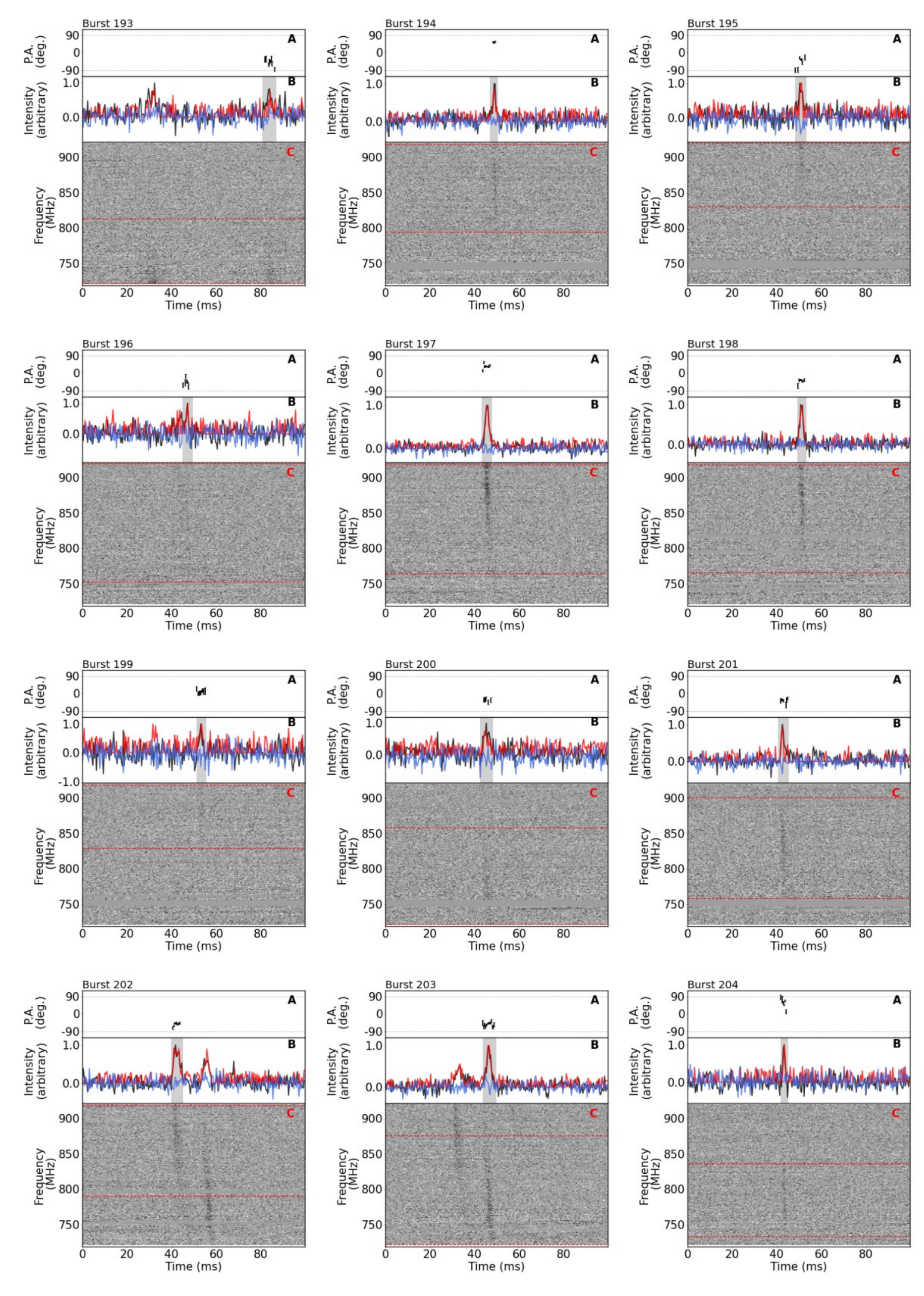}
\caption{}
\end{figure*}

\renewcommand{\thefigure}{A\arabic{figure} (Continued.)}
\addtocounter{figure}{-1}
\begin{figure*}
\centering
\includegraphics[width=.95\textwidth]{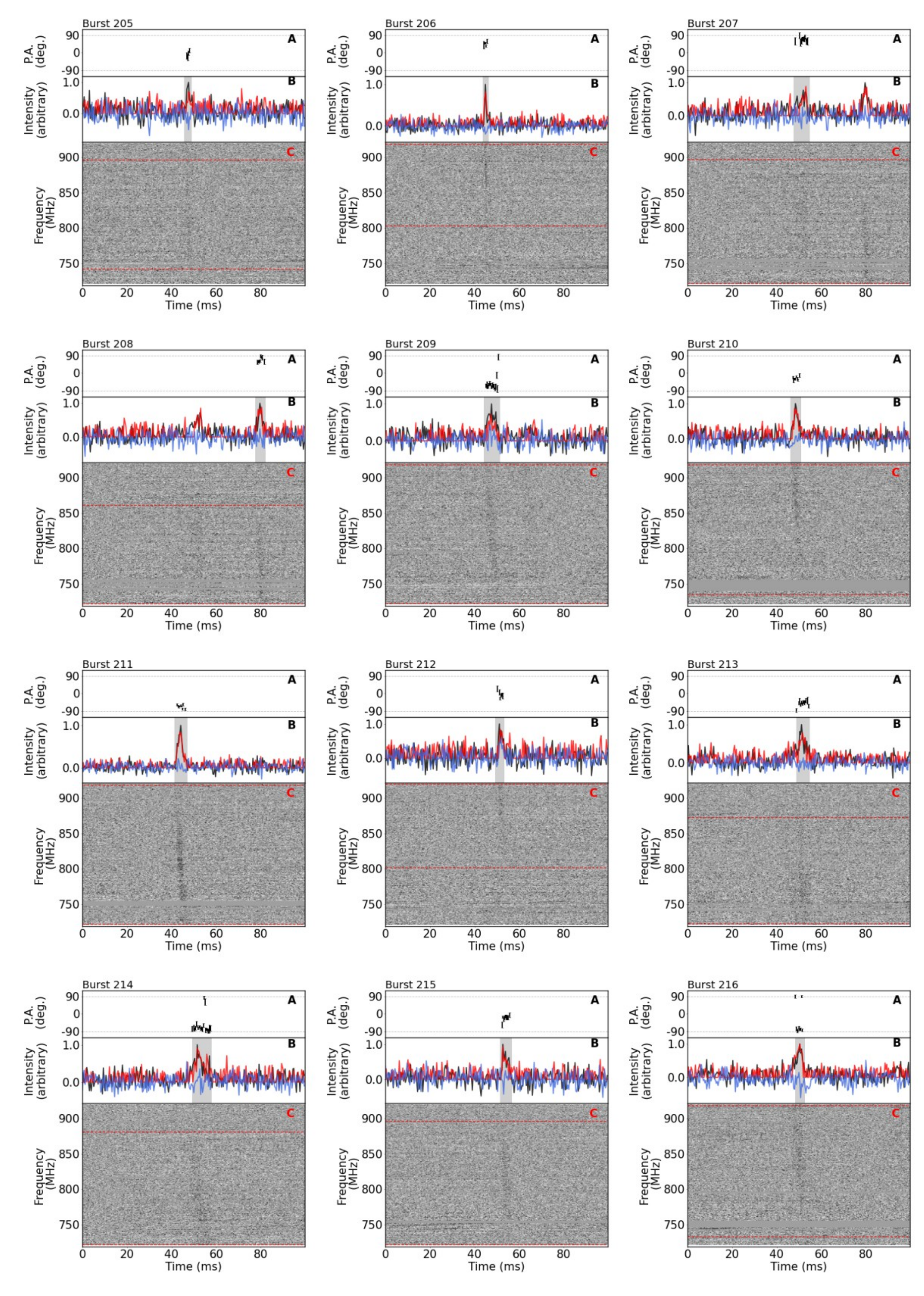}
\caption{}
\end{figure*}

\renewcommand{\thefigure}{A\arabic{figure} (Continued.)}
\addtocounter{figure}{-1}
\begin{figure*}
\centering
\includegraphics[width=0.95\textwidth]{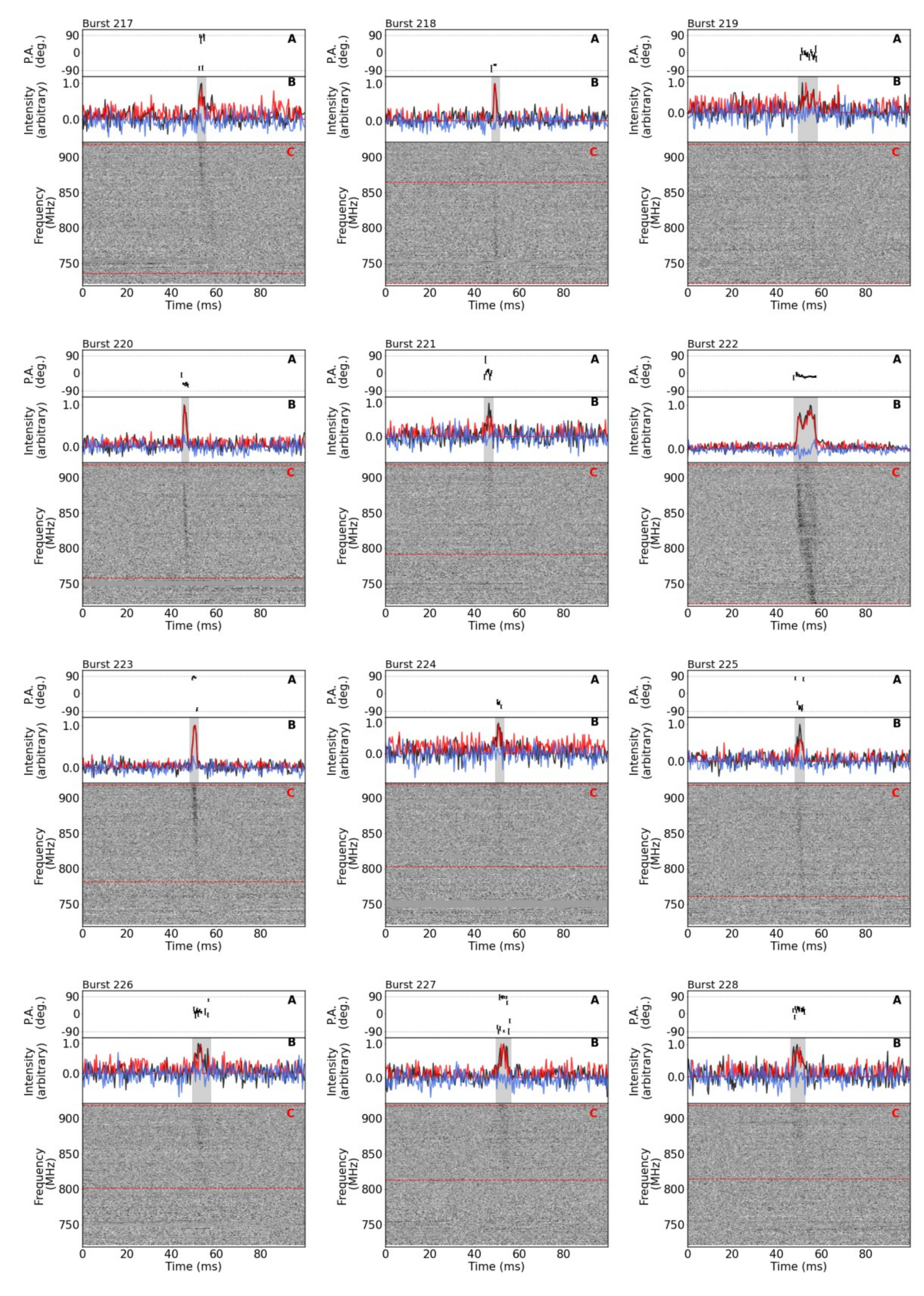}
\caption{}
\end{figure*}

\renewcommand{\thefigure}{A\arabic{figure} (Continued.)}
\addtocounter{figure}{-1}
\begin{figure*}
\centering
\includegraphics[width=0.95\textwidth]{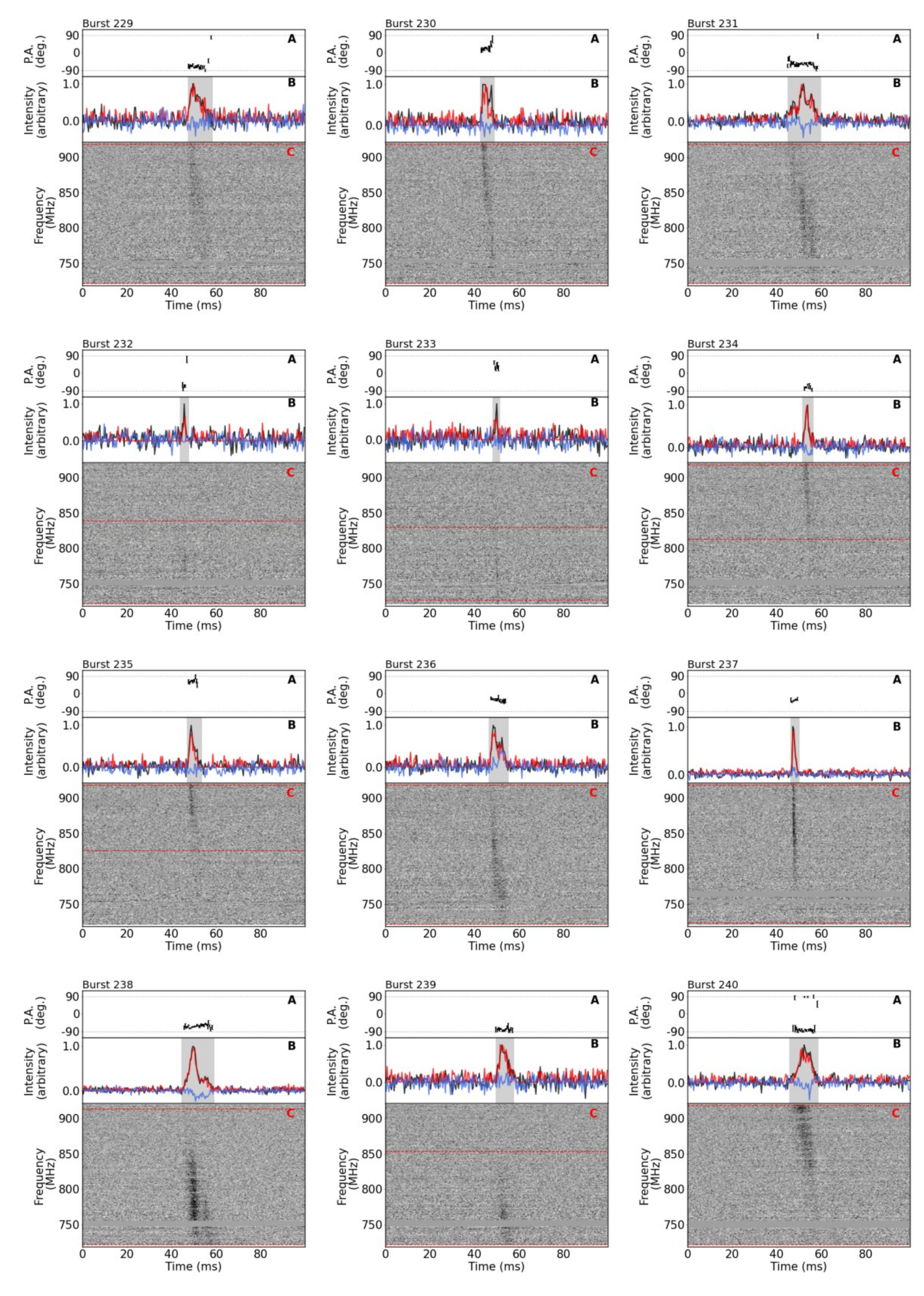}
\caption{}
\end{figure*}

\renewcommand{\thefigure}{A\arabic{figure} (Continued.)}
\addtocounter{figure}{-1}
\begin{figure*}
\centering
\includegraphics[width=0.95\textwidth]{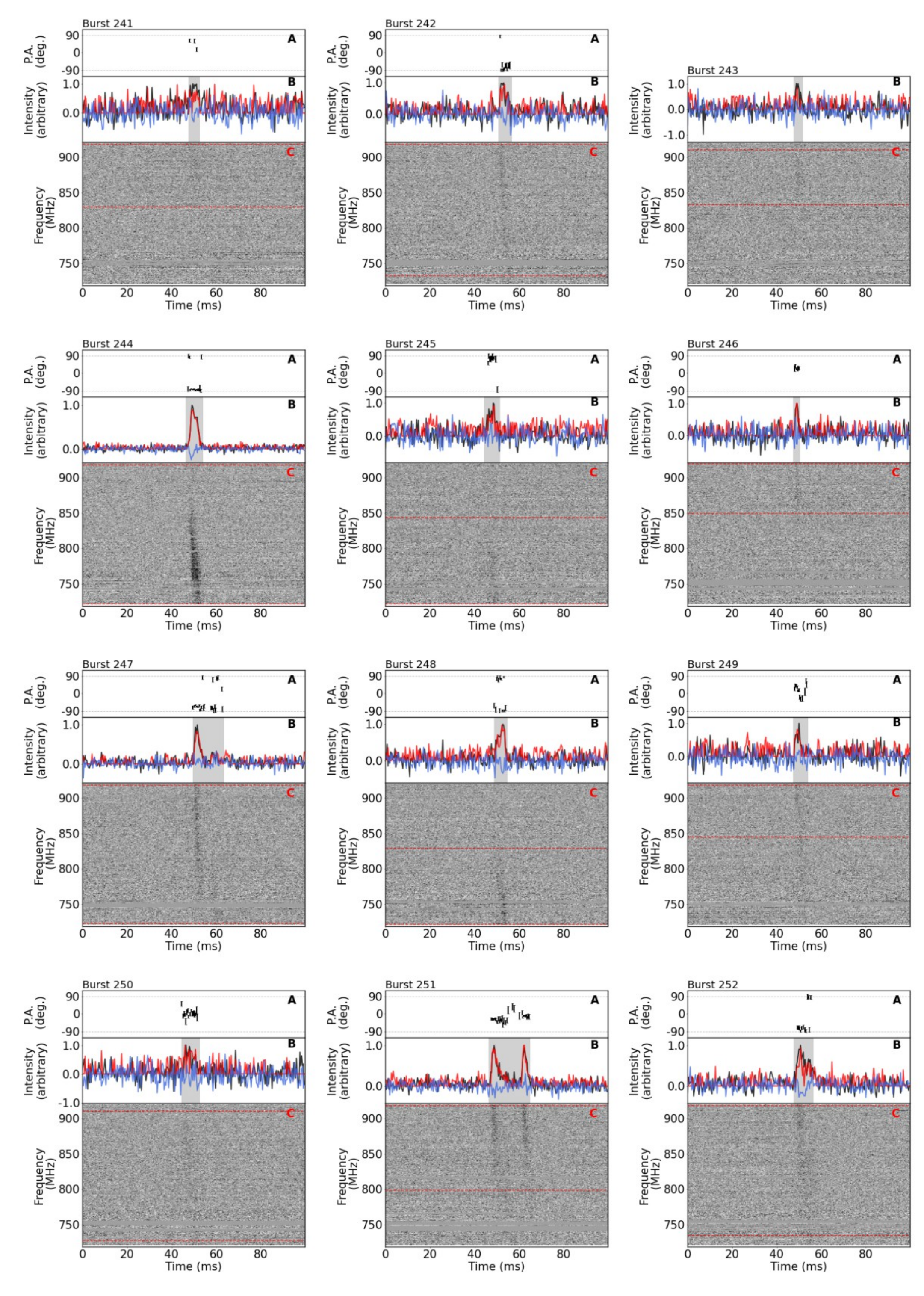}
\caption{}
\end{figure*}

\renewcommand{\thefigure}{A\arabic{figure} (Continued.)}
\addtocounter{figure}{-1}
\begin{figure*}
\centering
\includegraphics[width=0.95\textwidth]{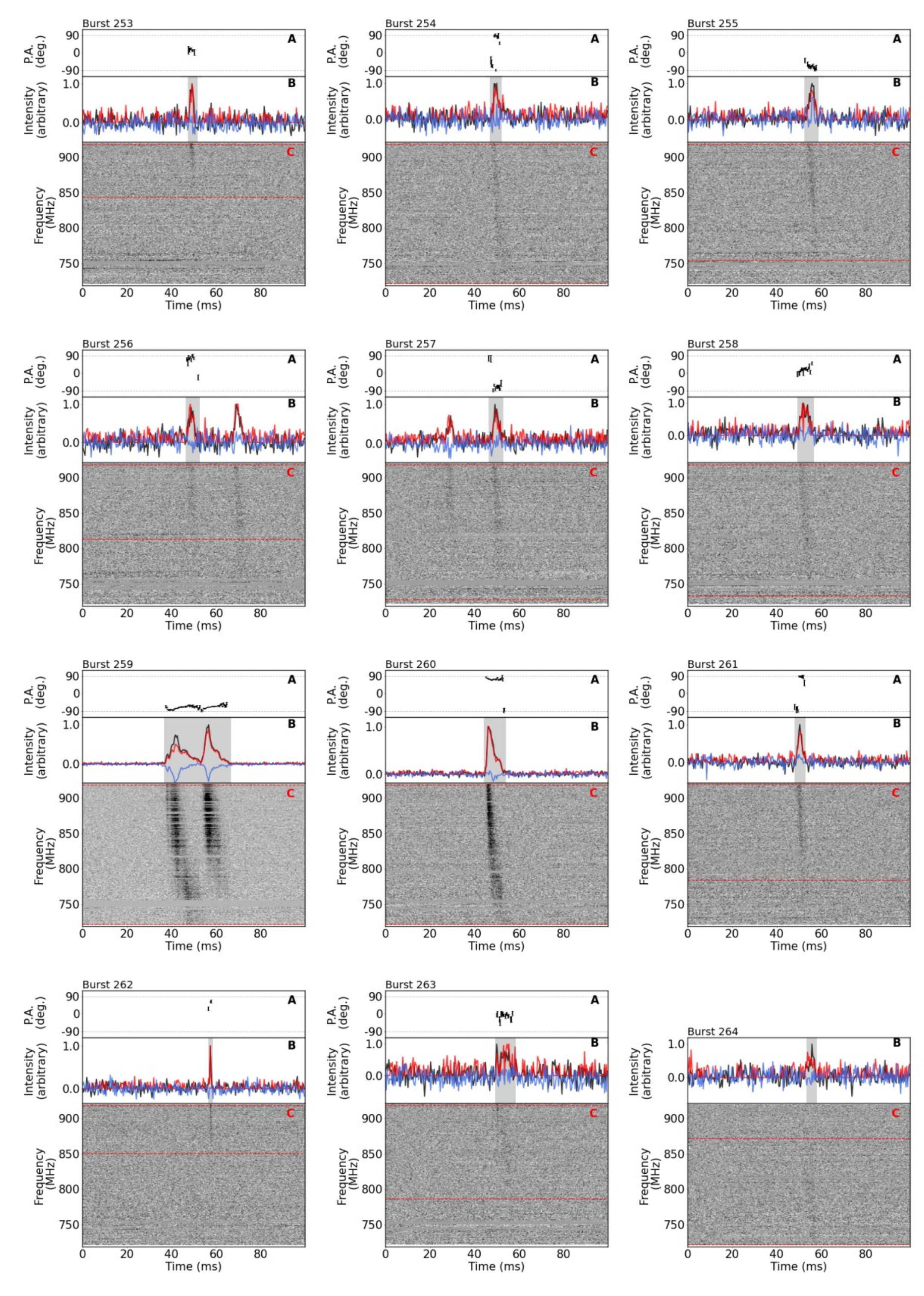}
\caption{}
\end{figure*}

\renewcommand{\thefigure}{A\arabic{figure} (Continued.)}
\addtocounter{figure}{-1}
\begin{figure*}
\centering
\includegraphics[width=0.95\textwidth]{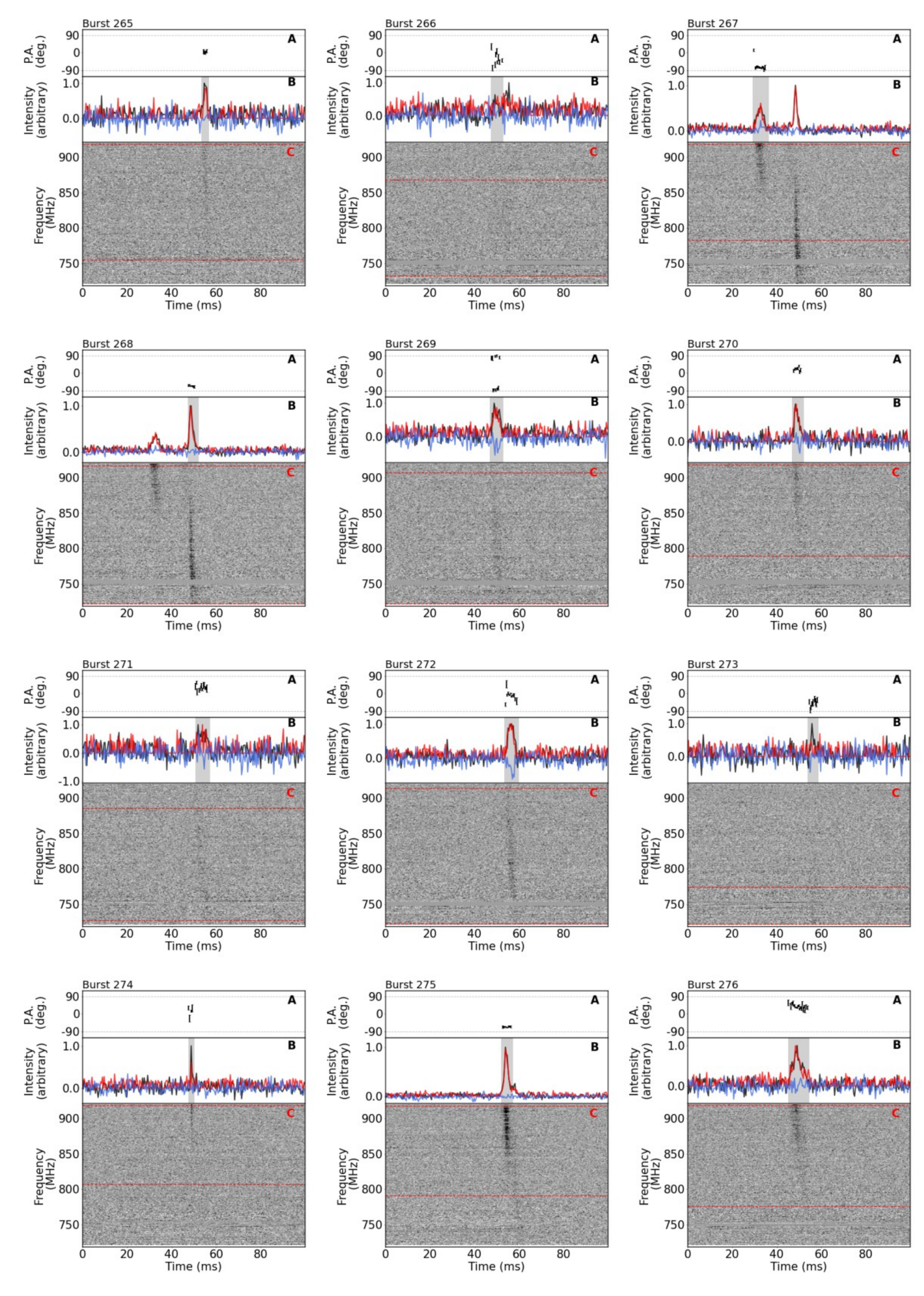}
\caption{}
\end{figure*}

\renewcommand{\thefigure}{A\arabic{figure} (Continued.)}
\addtocounter{figure}{-1}
\begin{figure*}
\centering
\includegraphics[width=0.95\textwidth]{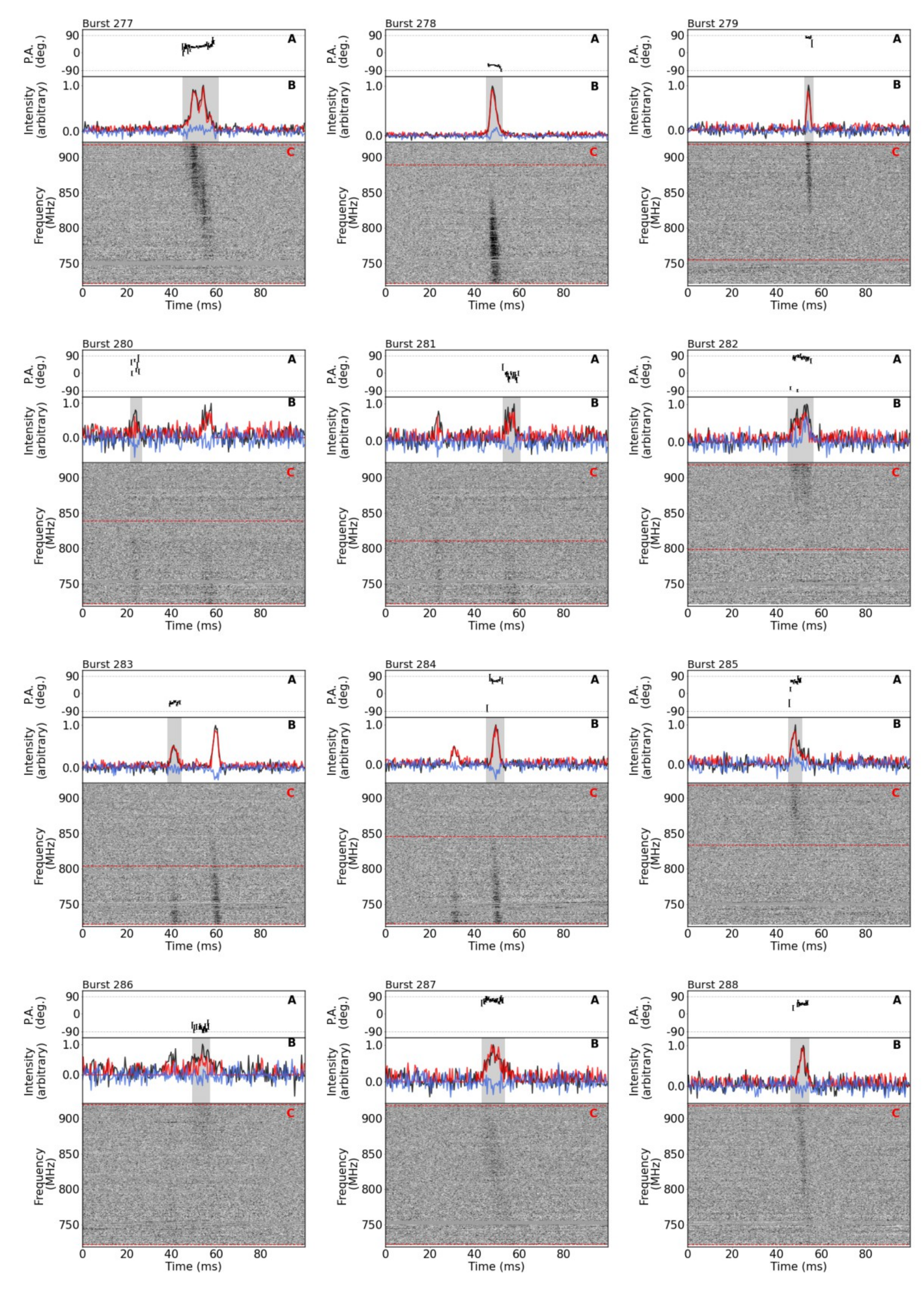}
\caption{}
\end{figure*}

\renewcommand{\thefigure}{A\arabic{figure} (Continued.)}
\addtocounter{figure}{-1}
\begin{figure*}
\centering
\includegraphics[width=0.95\textwidth]{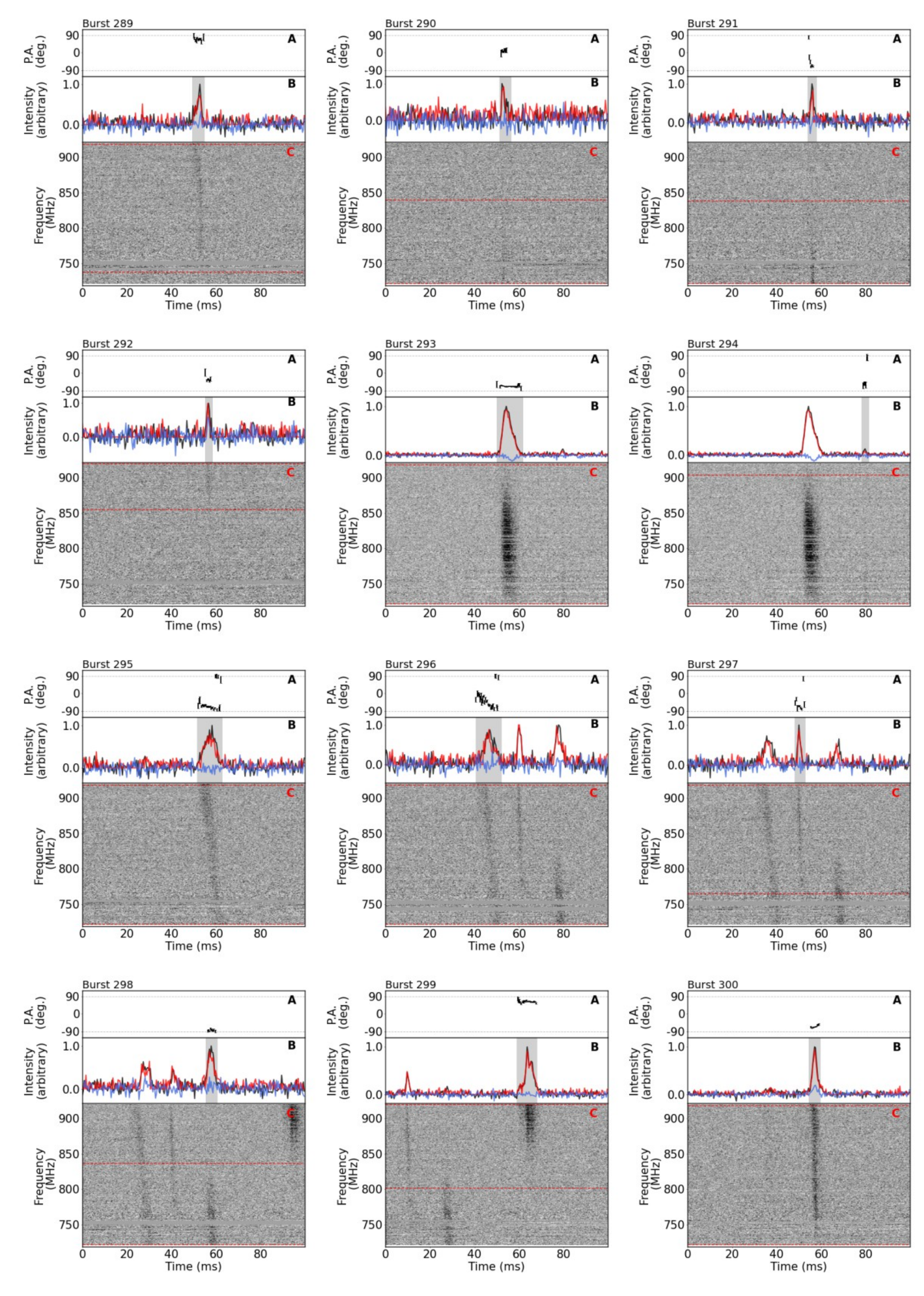}
\caption{}
\end{figure*}

\renewcommand{\thefigure}{A\arabic{figure} (Continued.)}
\addtocounter{figure}{-1}
\begin{figure*}
\centering
\includegraphics[width=0.95\textwidth]{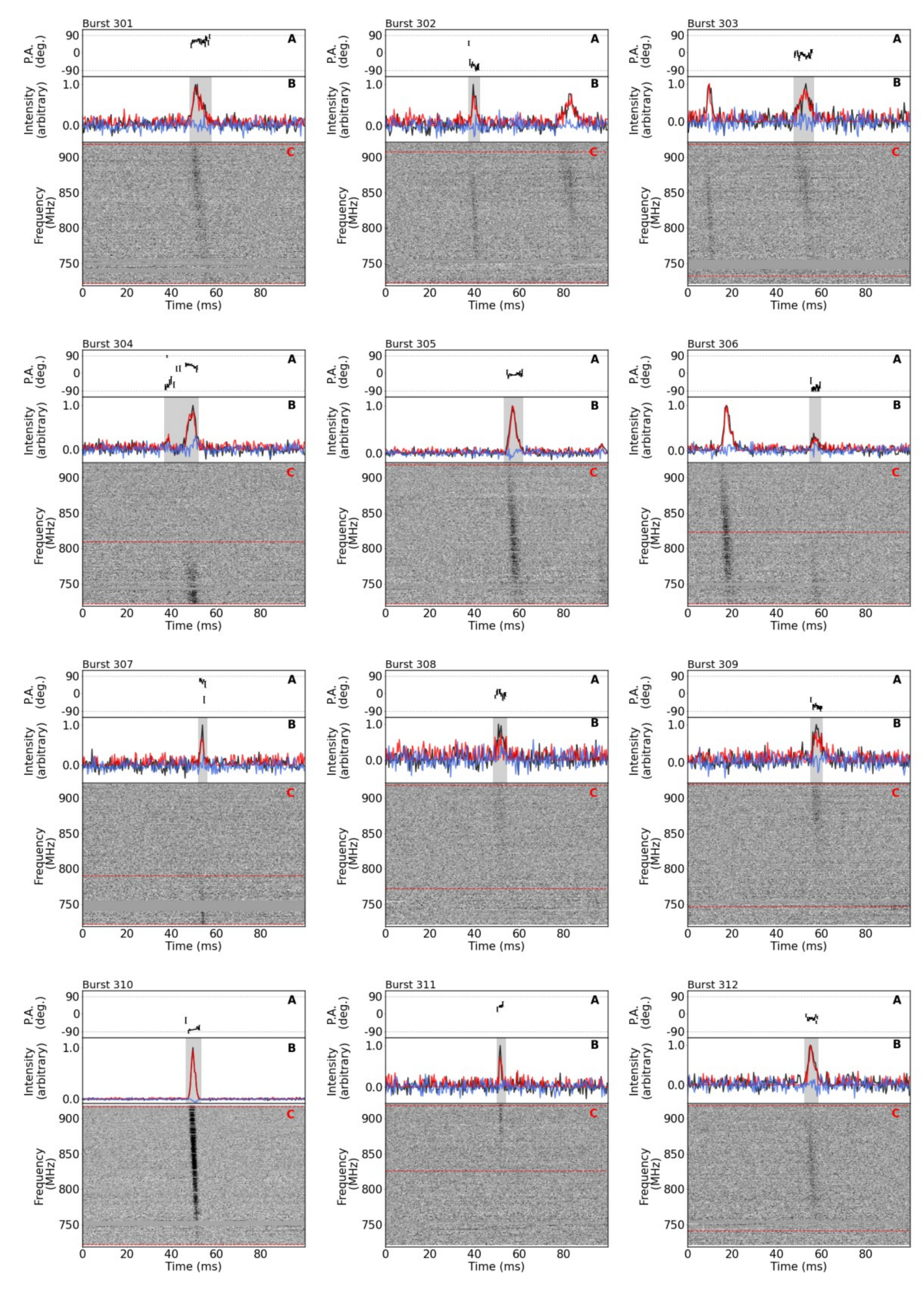}
\caption{}
\end{figure*}

\renewcommand{\thefigure}{A\arabic{figure} (Continued.)}
\addtocounter{figure}{-1}
\begin{figure*}
\centering
\includegraphics[width=0.95\textwidth]{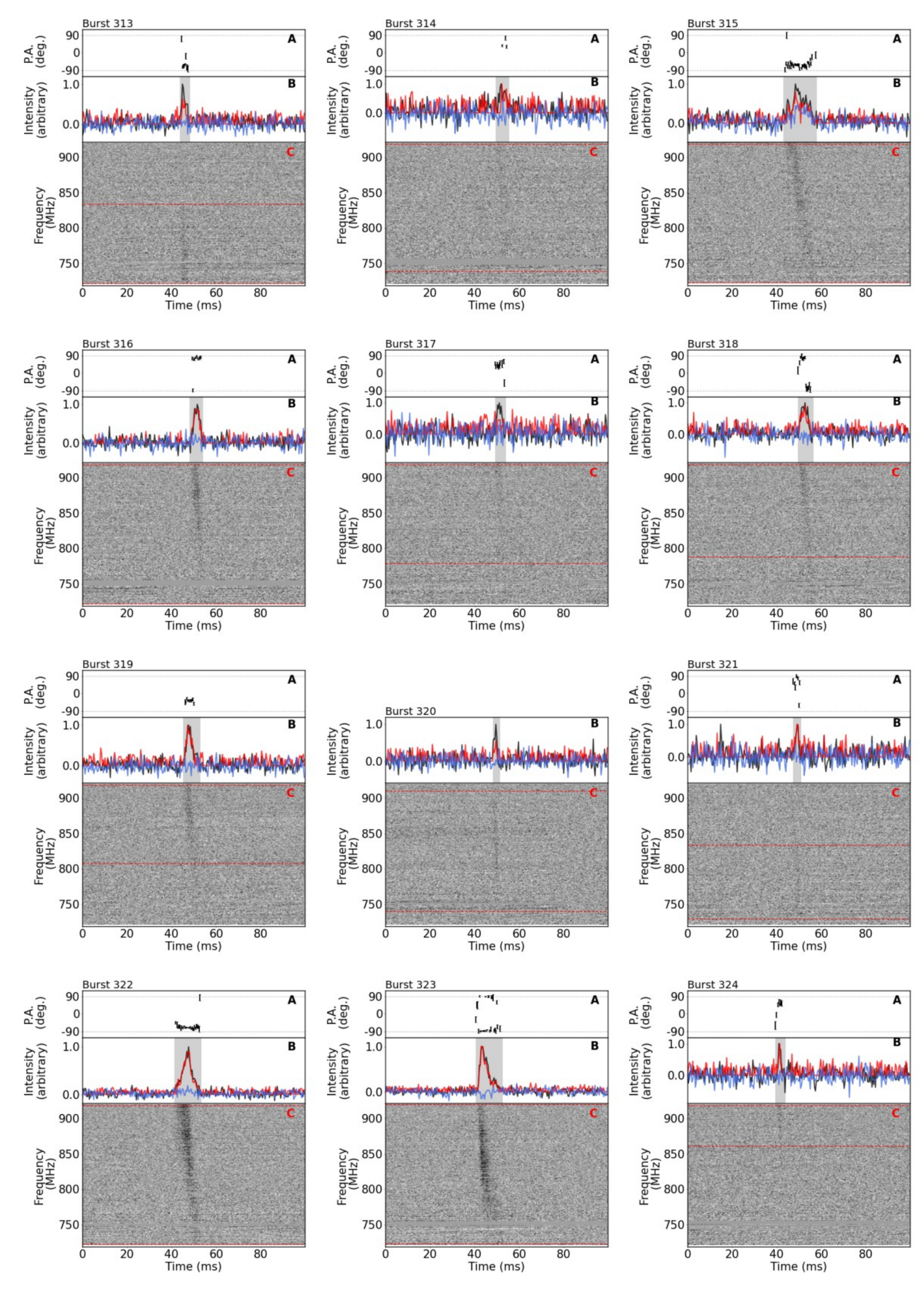}
\caption{}
\end{figure*}

\renewcommand{\thefigure}{A\arabic{figure} (Continued.)}
\addtocounter{figure}{-1}
\begin{figure*}
\centering
\includegraphics[width=0.95\textwidth]{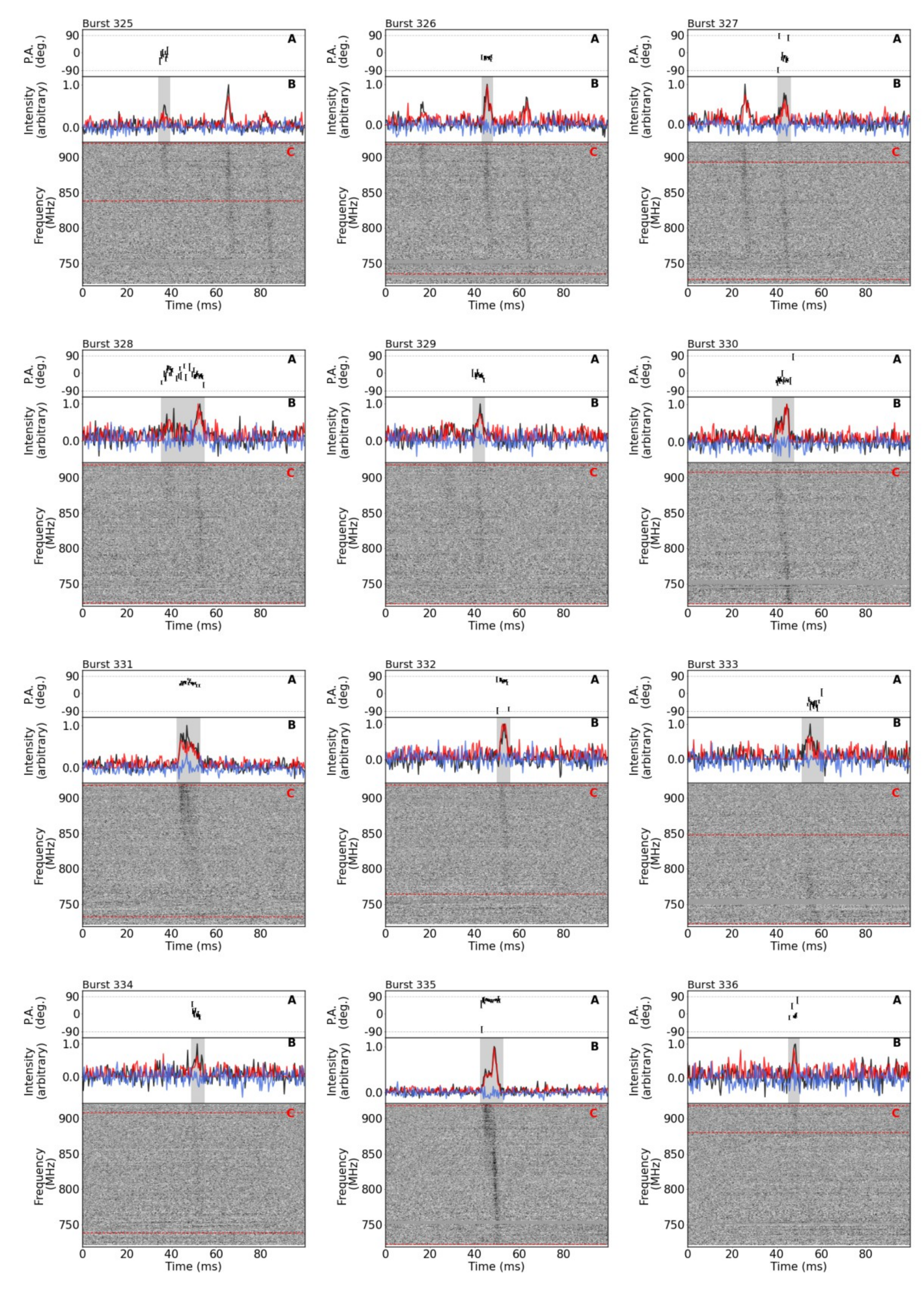}
\caption{}
\end{figure*}

\renewcommand{\thefigure}{A\arabic{figure} (Continued.)}
\addtocounter{figure}{-1}
\begin{figure*}
\centering
\includegraphics[width=0.95\textwidth]{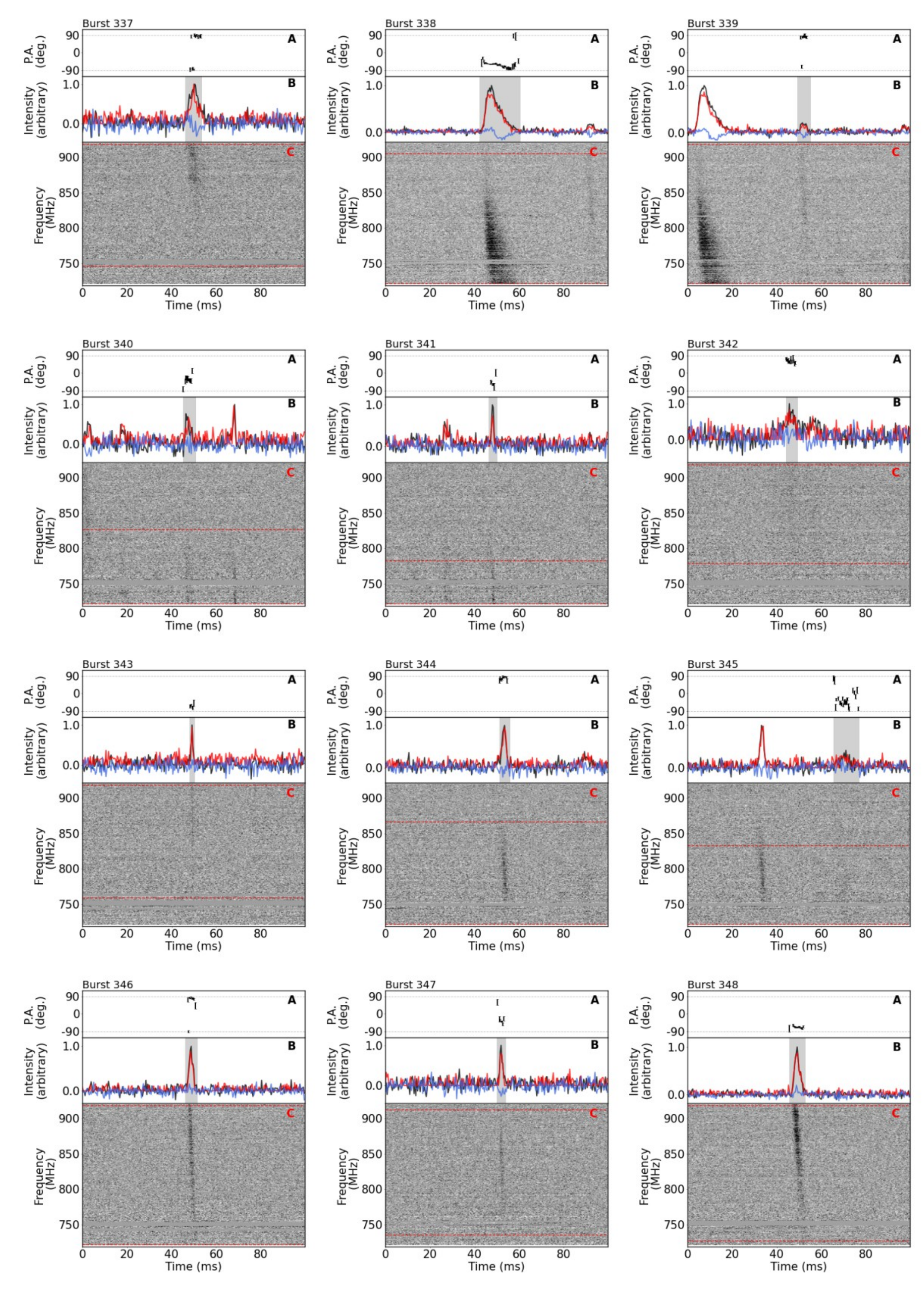}
\caption{}
\end{figure*}

\renewcommand{\thefigure}{A\arabic{figure} (Continued.)}
\addtocounter{figure}{-1}
\begin{figure*}
\centering
\includegraphics[width=0.95\textwidth]{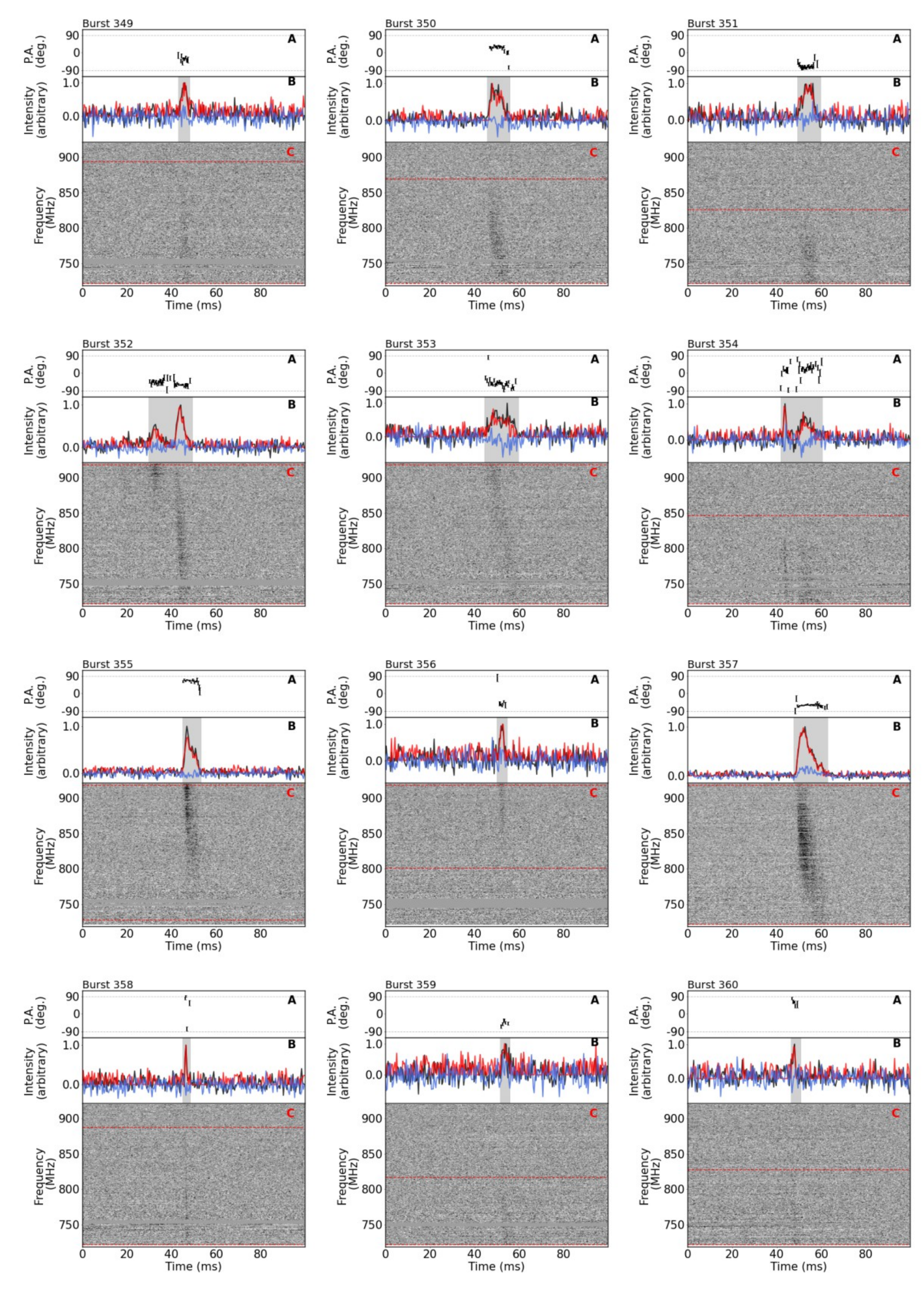}
\caption{}
\end{figure*}

\renewcommand{\thefigure}{A\arabic{figure} (Continued.)}
\addtocounter{figure}{-1}
\begin{figure*}
\centering
\includegraphics[width=0.95\textwidth]{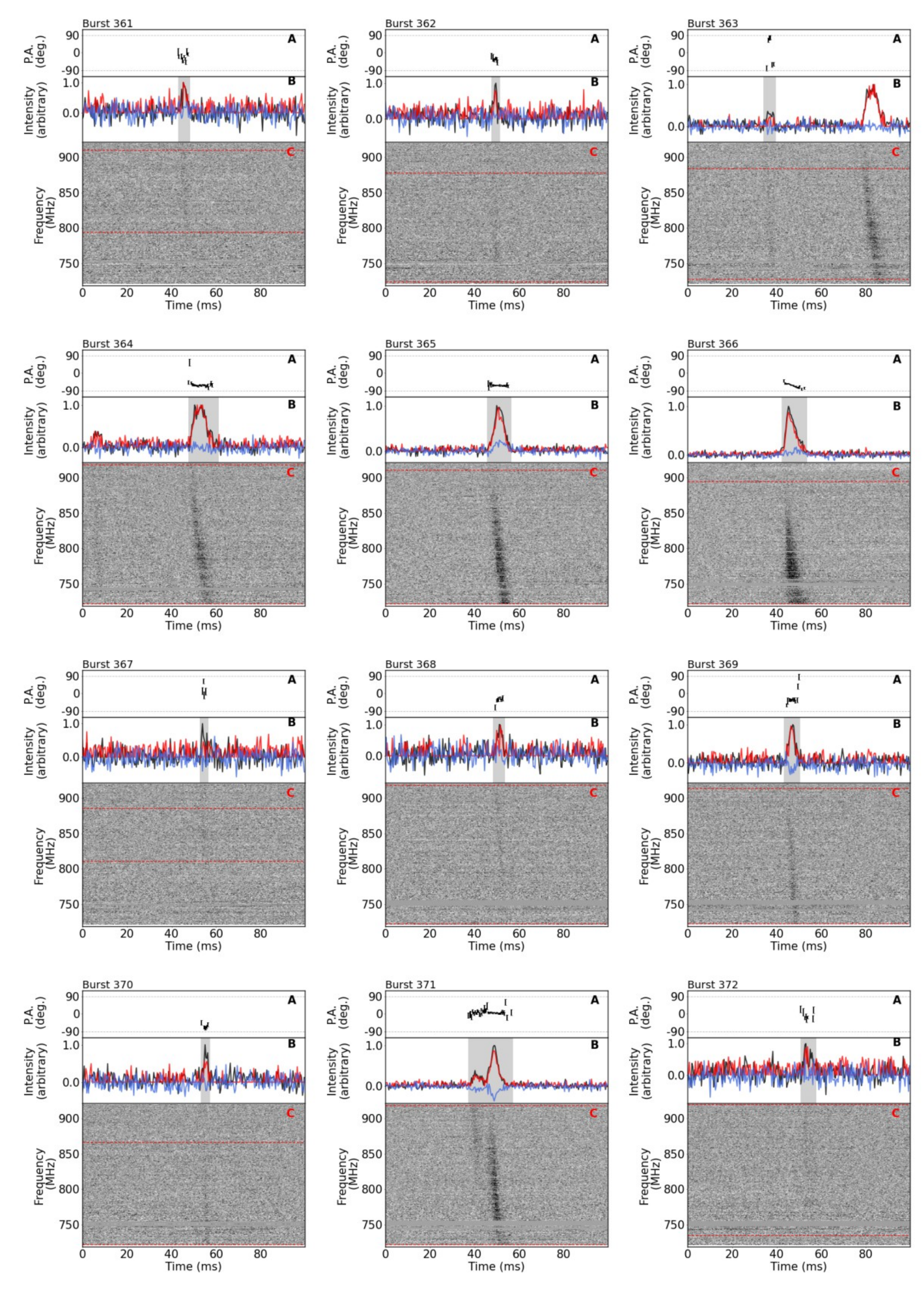}
\caption{}
\end{figure*}

\renewcommand{\thefigure}{A\arabic{figure} (Continued.)}
\addtocounter{figure}{-1}
\begin{figure*}
\centering
\includegraphics[width=0.95\textwidth]{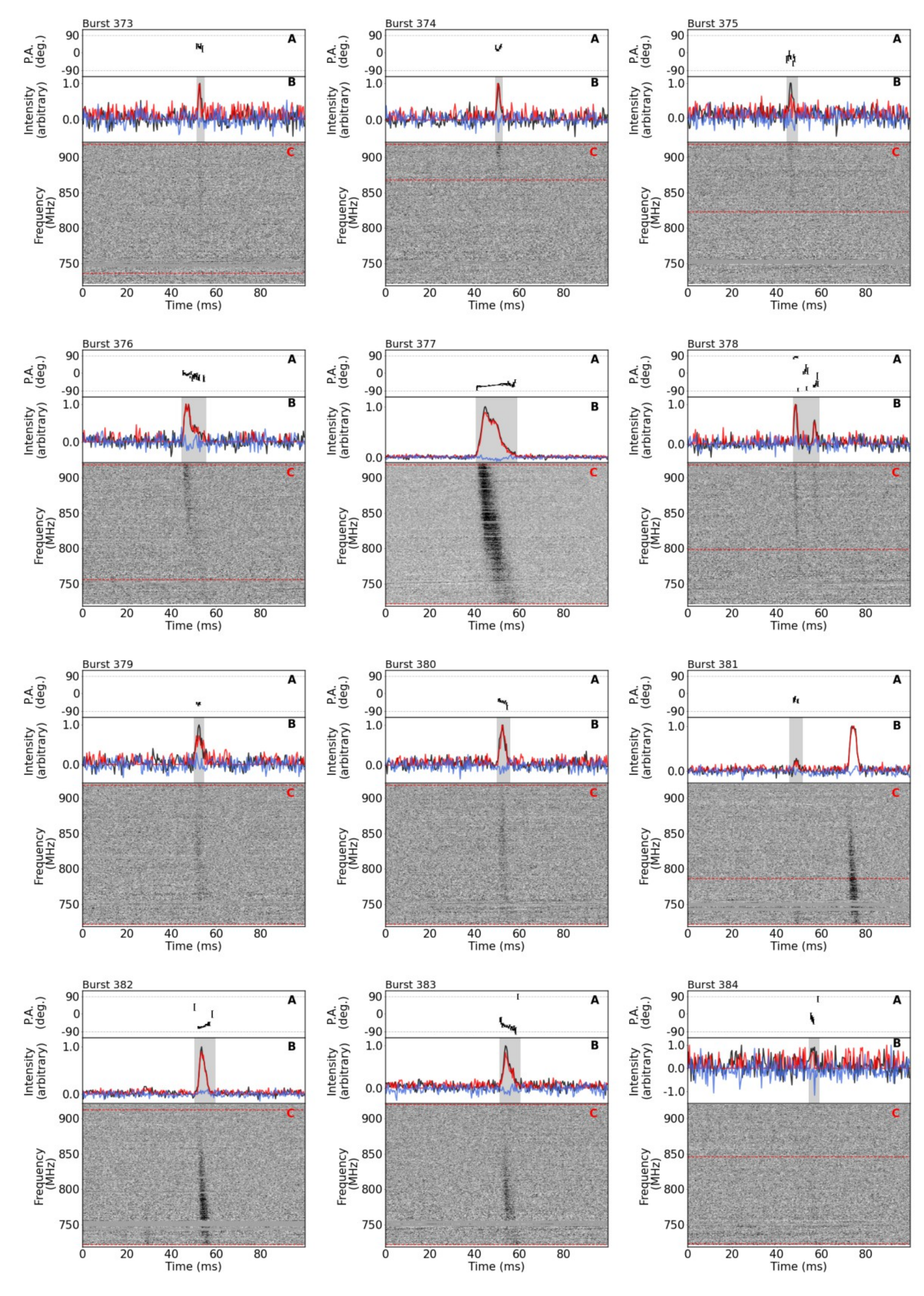}
\caption{}
\end{figure*}

\renewcommand{\thefigure}{A\arabic{figure} (Continued.)}
\addtocounter{figure}{-1}
\begin{figure*}
\centering
\includegraphics[width=0.95\textwidth]{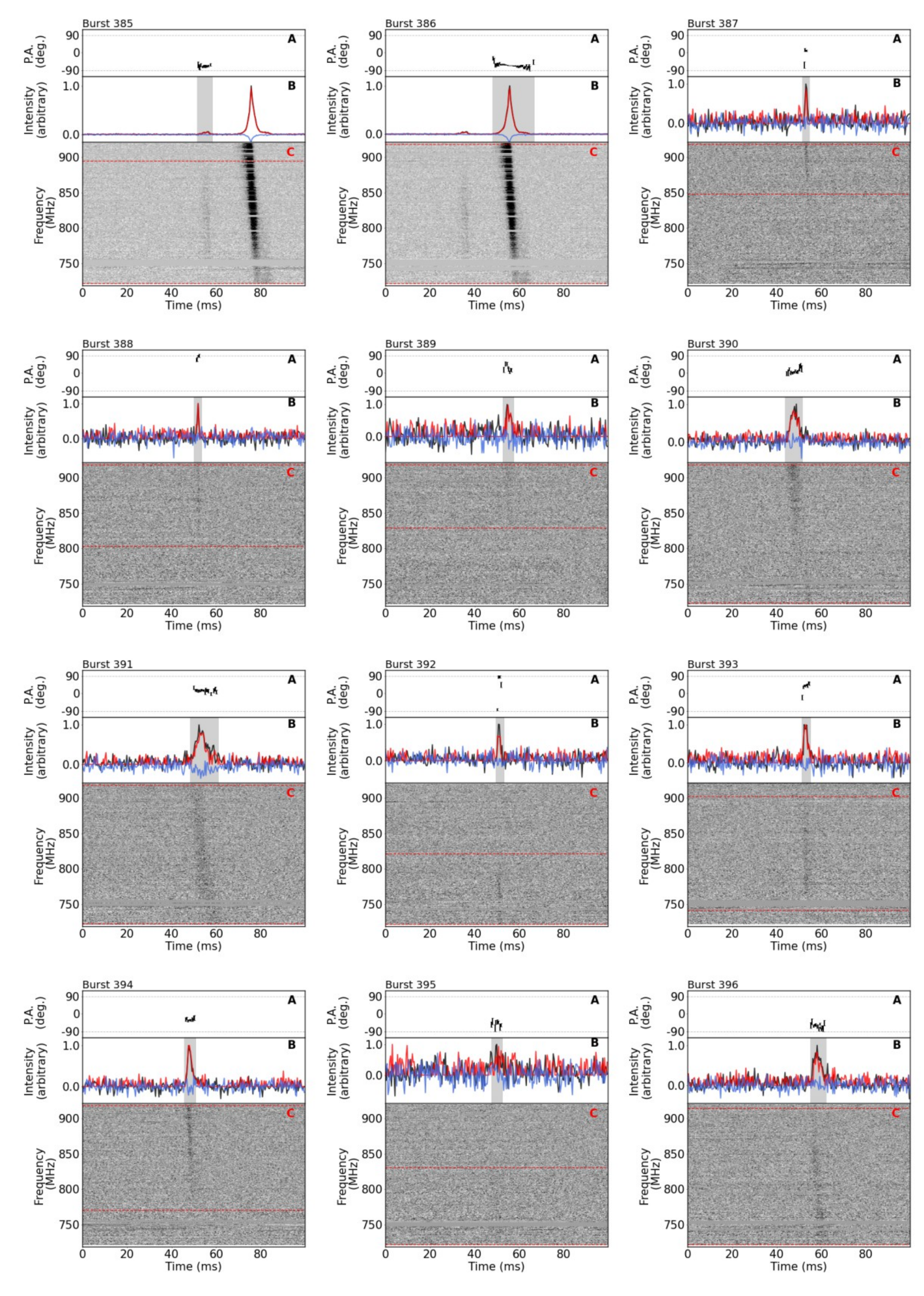}
\caption{}
\end{figure*}

\renewcommand{\thefigure}{A\arabic{figure} (Continued.)}
\addtocounter{figure}{-1}
\begin{figure*}
\centering
\includegraphics[width=0.95\textwidth]{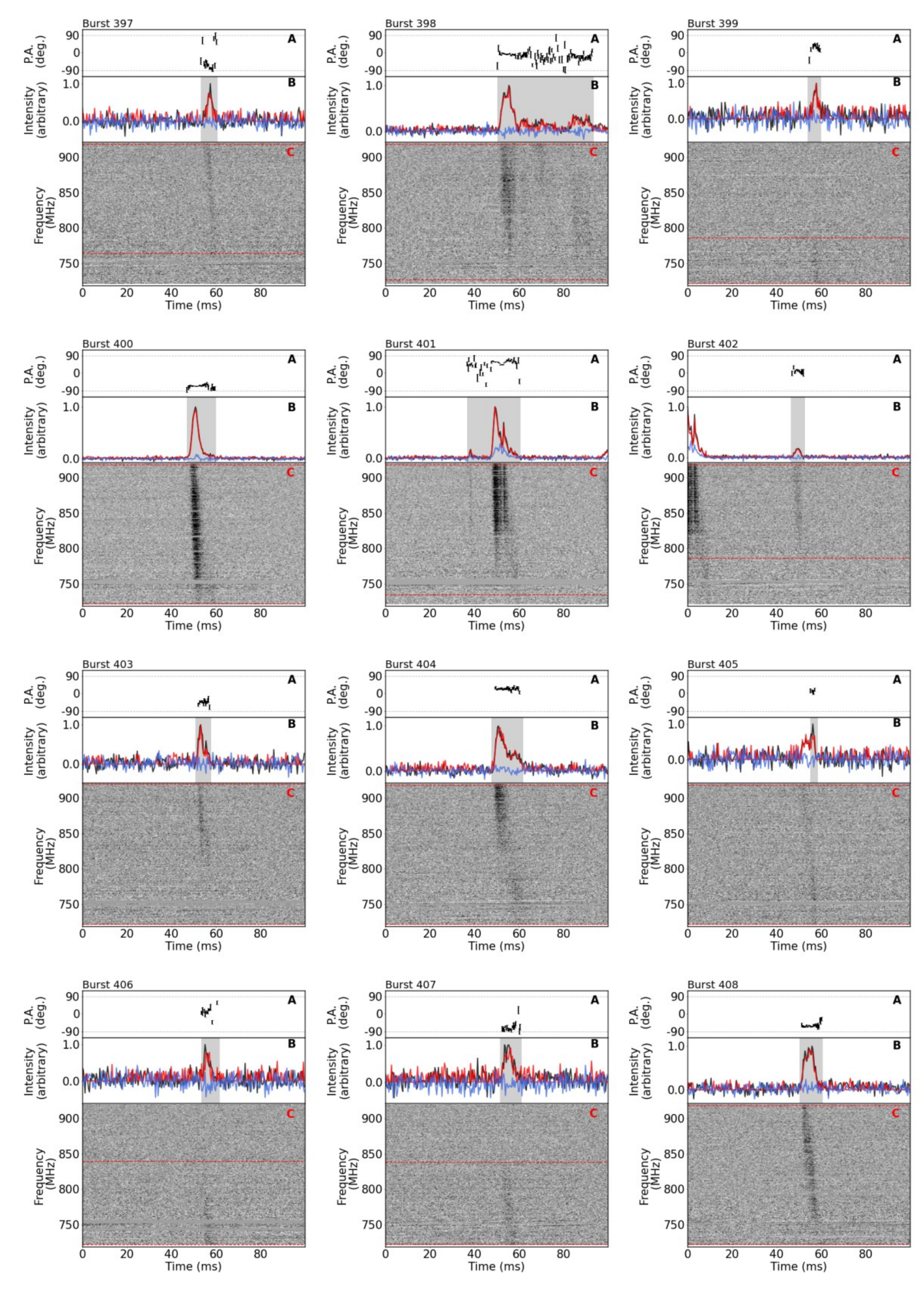}
\caption{}
\end{figure*}

\renewcommand{\thefigure}{A\arabic{figure} (Continued.)}
\addtocounter{figure}{-1}
\begin{figure*}
\centering
\includegraphics[width=0.95\textwidth]{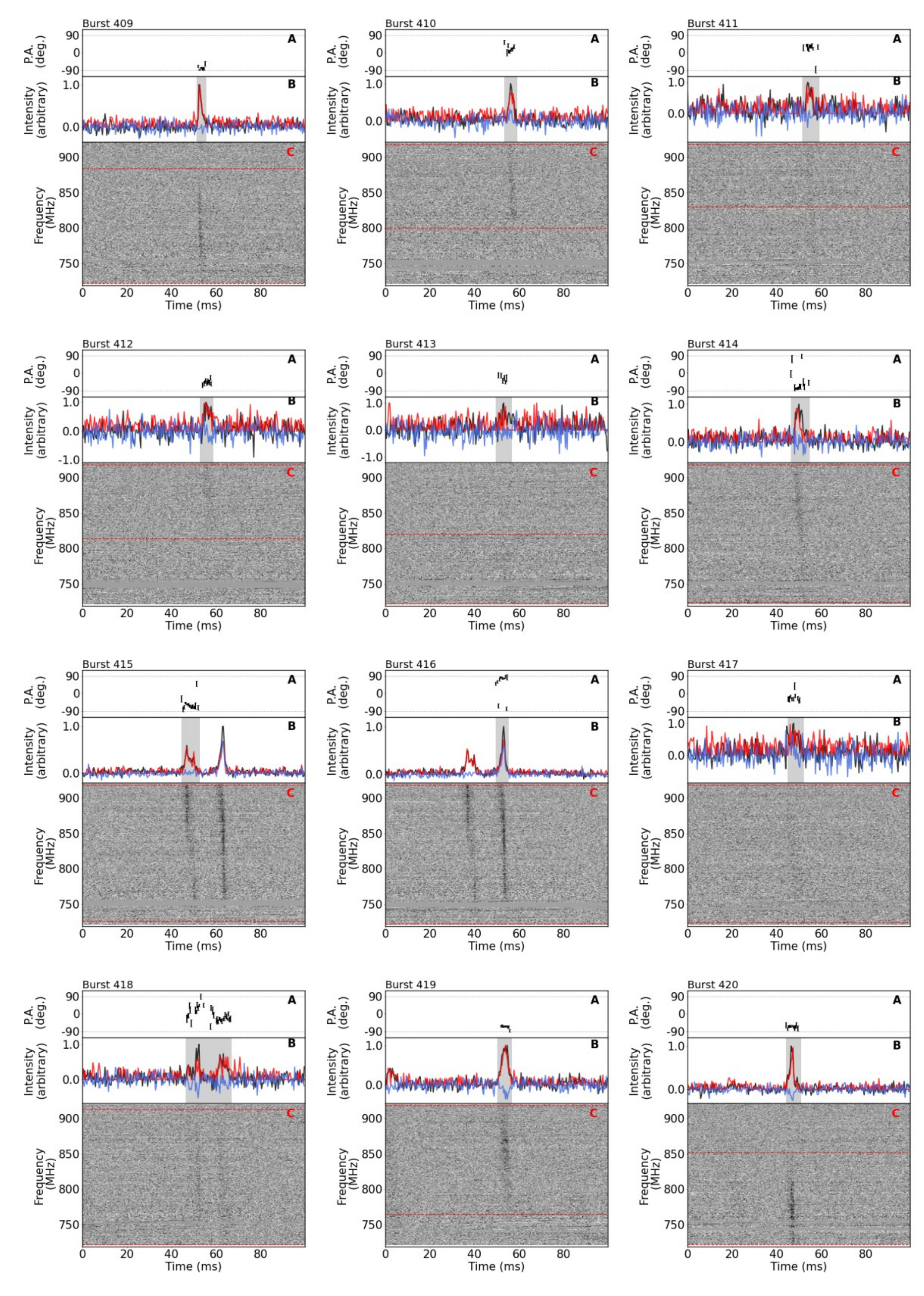}
\caption{}
\end{figure*}

\renewcommand{\thefigure}{A\arabic{figure} (Continued.)}
\addtocounter{figure}{-1}
\begin{figure*}
\centering
\includegraphics[width=0.95\textwidth]{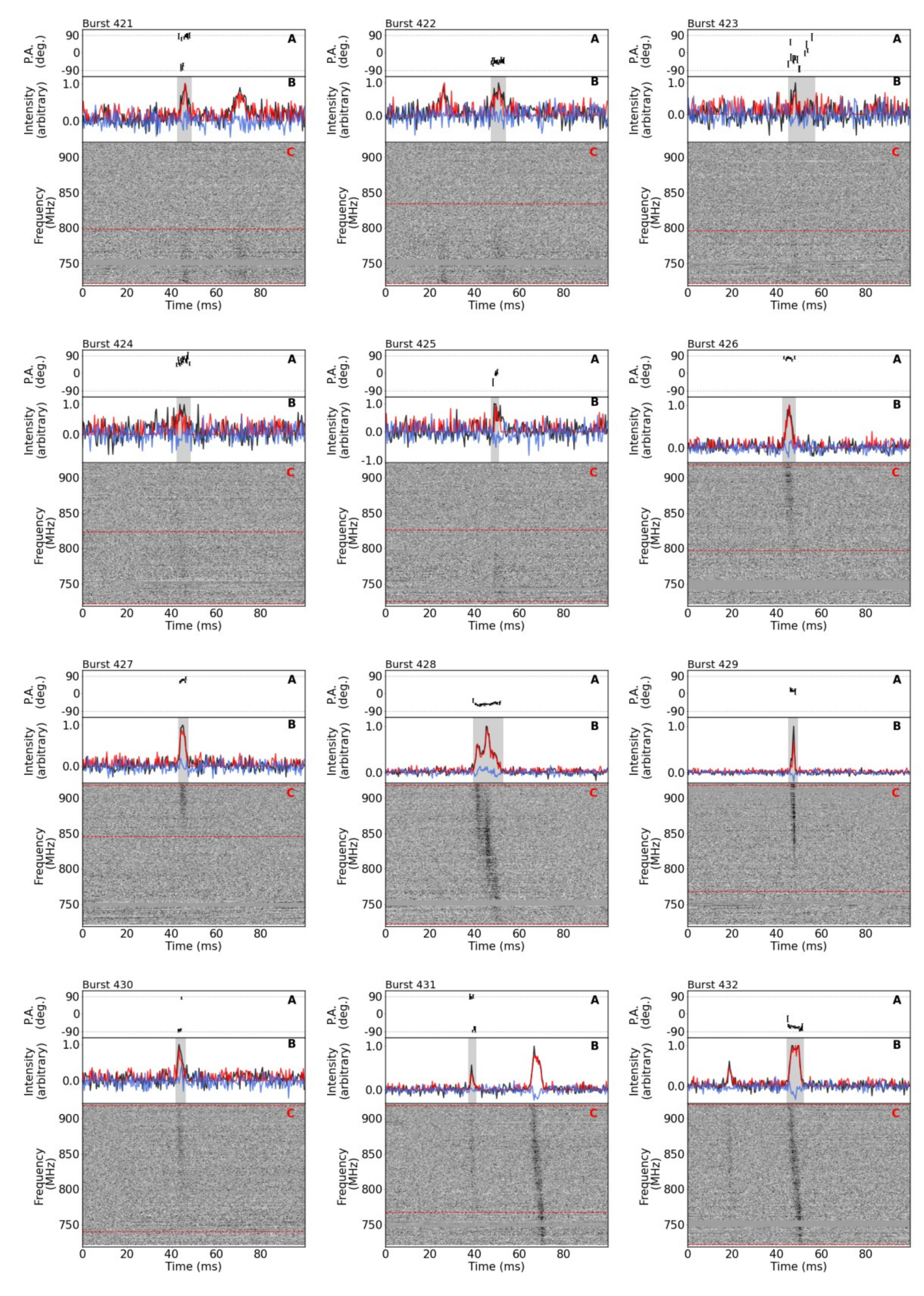}
\caption{}
\end{figure*}

\renewcommand{\thefigure}{A\arabic{figure} (Continued.)}
\addtocounter{figure}{-1}
\begin{figure*}
\centering
\includegraphics[width=0.95\textwidth]{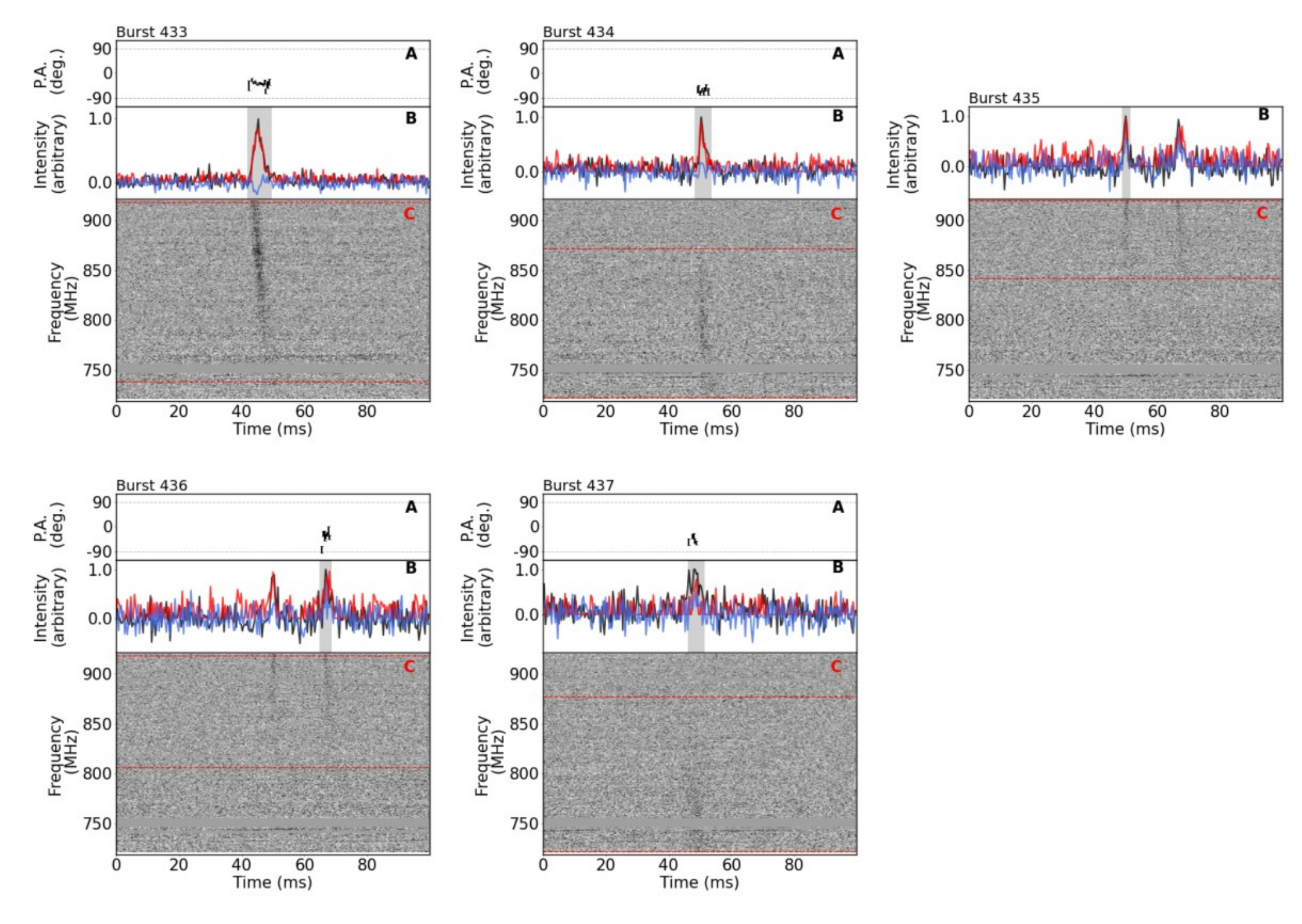}
\caption{}
\end{figure*}



\end{document}